\documentclass[usenatbib]{mn2e}
\usepackage{apjfonts}
\usepackage{epsfig}
\usepackage{amssymb}
\usepackage{amsmath}
\usepackage{fixltx2e} 
\usepackage{url}

\newcommand{\scaleup}{}

\newcommand{\apjl}{ApJL}
\newcommand{\apjs}{ApJS}

\newcommand{\aap}{A\&A}

\newcommand{\mergercalcurl}{\url{http://www.cfa.harvard.edu/~phopkins/Site/mergercalc.html}}
\newcommand\plotone[1]
 {\centering \leavevmode \includegraphics[width={0.99\columnwidth}]{#1}}
\newcommand{\plotside}[1]
 {\centering \leavevmode \includegraphics[width={0.95\textwidth}]{#1}}
\newcommand{\acknowledgments}{\begin{small}\section*{Acknowledgments}\end{small}}
\newcommand\altaffilmark[1]{$^{#1}$}
\newcommand\altaffiltext[1]{$^{#1}$}
\newcommand{\etal}{et al.}

\newcommand{\msun}{M_{\sun}}

\newcommand{\stabilityparam}{\chi_{\rm eff}}

\title[``Isolated'' Galaxies and Secular Evolution]{When Should We Treat Galaxies 
as Isolated?}

\author[Hopkins \etal]{
\parbox[t]{\textwidth}{ 
Philip F. Hopkins\altaffilmark{1},\thanks{E-mail:phopkins@astro.berkeley.edu} 
Du{\v s}an Kere{\v s}\altaffilmark{2},
Chung-Pei Ma\altaffilmark{1},
\&\ Eliot Quataert\altaffilmark{1} }
\vspace*{6pt} \\
\altaffiltext{1}{Department of Astronomy, University of California 
Berkeley, Berkeley, CA 94720} \\
\altaffiltext{2}{Harvard-Smithsonian Center for Astrophysics, 
60 Garden Street, Cambridge, MA 02138}}

\date{Submitted to MNRAS, February 12, 2009}
\begin{document}
\maketitle
\label{firstpage}

\begin{abstract}

Traditionally, secular evolution is defined as evolution of systems where 
the internal growth of structure and instabilities dominates the  
growth via external drivers (e.g.\ accretion and mergers). 
Most study has focused on ``isolated'' galaxies, where seed asymmetries may 
represent realistic cosmological substructure, but subsequent evolution 
ignores galaxy growth and interactions. Large-scale modes in the disk 
then grow on a timescale of order a disk rotation period ($\sim0.1-1$\,Gyr). If, however, 
galaxies evolve cosmologically on a shorter timescale, then it may not 
be appropriate to consider them ``isolated.'' 
We outline simple scalings to ask whether, 
under realistic conditions, the timescale for secular evolution is shorter than 
the timescale for cosmological accretion and mergers. We show that this is the case 
in a relatively narrow, but important range of perturbation 
amplitudes corresponding to substructure or mode/bar fractional amplitudes $\delta\sim0.01-0.1$, 
the range of most interest for observed strong bars and most pseudobulges. 
At smaller amplitudes $\delta\ll0.1$, systems are not isolated: typical disks will grow 
by accretion at a comparable level over even a single dynamical time. 
At larger amplitudes $\delta\gg0.1$, the 
evolution is no longer secular; the direct gravitational evolution of the 
seed substructure swamps the internal disk response.
We derive criteria for when disks can be well-approximated as ``isolated'' as a 
function of mass, redshift, and disk stability. 
The relevant parameter space shrinks at higher mass, higher disk stability,
and higher-$z$ as accretion rates increase. 
The cosmological rate of galaxy evolution 
also defines a maximum bar/mode lifetime of practical interest, 
of $\sim0.1\,t_{\rm Hubble}(z)$. Longer-lived modes 
will encounter cosmological effects and will de-couple from their 
drivers (if they are driven). 

\end{abstract}


\section{Introduction}
\label{sec:intro}

Isolated disk galaxies are prone to a number of important instabilities 
that play a major role in shaping observed late-type disk and bulge populations, 
with the most well-known and well-studied being 
the traditional bar and spiral instabilities. 
Both bars and spiral structure are ubiquitous in the 
local disk population \citep{marinova:bar.frac.vs.freq,
menendez:2mass.bars,barazza:bar.colors}, and 
their abundance appears comparable 
at higher redshifts \citep{sheth:bar.frac.evol,sheth:bar.evol.cosmos,
jogee:bar.frac.evol}. 
By amplifying small perturbations into coherent,  
long-lived, 
large-scale non-axisymmetric modes, these structures 
enable disks to evolve significantly -- re-distributing 
material in angular momentum and phase space -- in a few orbital periods. 
As a consequence, observations and simulations indicate that these structures are 
important in shaping the cosmological evolution of 
disk sizes, scale heights, and 
the abundance, structural properties, and 
mass fraction in ``pseudobulges'' (disk-like 
bulges that result from angular momentum exchange in these modes), 
a population increasingly 
prominent in low-mass and later-type disk galaxies 
\citep[e.g.][]{debattista:pseudobulges.a,
kormendy.kennicutt:pseudobulge.review,weinzirl:b.t.dist}. 

Traditionally, the growth and evolution of 
these {\em global} modes is referred to as ``secular'' evolution: by definition, 
evolution that is slow relative to the local dynamical time. 
This contrasts with violent relaxation -- seen in e.g.\ galaxy-galaxy 
major mergers -- in which the potential fluctuates 
on short timescales, and {\em local} instabilities, involving e.g.\ clumping, 
star formation, and formation of bars on small scales (sub-kpc). 

As a consequence, the secular 
evolutionary channel has, for the most part, been 
studied in the context of isolated galaxies. 
Given an isolated, self-gravitating stellar (or stellar+gas) 
disk that meets certain instability criteria, small 
non-zero amplitude in the large-scale modes that identify morphological 
bar and spiral patterns 
(characteristic wavelength 
of order the disk length) will grow exponentially on a timescale 
of a few orbital periods \citep[see e.g.\ the discussion in][]{binneytremaine}. 
The evolution and dynamics of these modes have been 
well-studied in idealized cases of isolated disks with properties 
similar to the Milky Way, but by design bar or spiral wave-unstable 
\citep{schwarz:disk-bar,
athanassoula:bar.orbits,
pfenniger:bar.dynamics,
weinberg:bar.dynfric,
combes:pseudobulges,
hernquistweinberg92,
friedli:gas.stellar.bar.evol,
athanassoula:bar.halo.growth,
athanassoula:bar.vs.concentration,
athanassoula:bar.evol.in.int,
weinberg:bar.res.requirements,
kaufmann:gas.bar.evol,
patsis:gas.flow.in.bars,
mayer:lsb.disk.bars,
berentzen:bar.destruction.in.int,
berentzen:self.damping.tidal.bar.generation,
foyle:two.component.disk.evol.from.bars}.
In particular this informed 
study of the role of secular evolution in shaping 
galaxy sizes, dynamics, and morphology. 

However, in $\Lambda$CDM cosmologies, 
structure grows via continuous accretion and mergers. 
Although major mergers are rare, 
both theoretical calculations and observations suggest that 
minor mergers are ubiquitous, and 
accretion of new cold gas is rapid in low-mass galaxies 
\citep{woods:tidal.triggering,
maller:sph.merger.rates,barton:triggered.sf,
woods:minor.mergers,stewart:mw.minor.accretion}. 
Together with the typical substructure present in $\Lambda$CDM halos 
\citep{taylor:substructure.evolution,gao:subhalo.mf}, this suggests 
the concern that there may not be in practice such a thing as 
an ``isolated'' galaxy at the level of interest.

More recent studies of secular evolution have 
therefore focused on more realistic 
scenarios, exploiting merger histories from cosmological simulations 
in semi-idealized studies of single galaxies 
\citep{bournaud:gas.bar.renewal.destruction,
benson:heating.model,gauthier:triggered.bar.from.substructure,
berentzen:disk.bar.in.live.halos,kaufmann:gas.bar.evol,
curir:bar.in.cosmo.halos,kazantzidis:mw.merger.hist.sim,
romanodiaz:substructure.driven.bars}. These simulations again 
reveal bars and spiral structure to be prominent -- 
arguably more so than in isolated simulations -- but it is less 
clear whether their formation and evoluton can be attributed 
to the same secular processes at work in isolated systems, 
or whether they are driven systems owing to 
substructure and accretion in the galaxy disk and halo. 

The important question for models is: can any 
galaxy in a realistic cosmological context 
still be approximated as ``isolated'' for 
certain purposes? If so, in what regimes as a function of redshift, galaxy mass, 
and internal properties is this applicable? 
What are the corresponding implications for interpretation of 
bar fractions and lifetimes? And ultimately, what does this imply for the  
importance of isolated secular evolution in driving 
the evolution of galaxies and formation of bulges? 

In this paper, we attempt to address these questions by means of 
a simple comparison of cosmological 
accretion rates and 
characteristic timescales for secular evolution. This approach allows us to 
identify the regimes where galaxies can be safely considered 
``isolated'' versus where cosmological effects may not be 
negligible. We show that there is an interesting regime of secular 
modes with fractional mass/amplitude $\sim0.1$ where the 
secular growth mode dominates and the isolated galaxy approximation 
is good (\S~\ref{sec:cosmo}). We show how this scales with 
galaxy mass, redshift, and disk stability properties (\S~\ref{sec:mz}), 
and identify some basic consequences for the 
lifetimes of large-scale modes in disks (\S~\ref{sec:lifetimes}). 
Our goal is not a definitive description of secular evolution, 
but rather to provide a set of simple initial constraints to provide 
context for more detailed studies of the interesting parameter space. 

Throughout, we adopt an 
$\Omega_{M}=0.3$, $\Omega_{\Lambda}=0.7$, $h=0.7$ 
cosmology, but our conclusions are not sensitive to 
the choice within the range allowed by 
present observations 
\citep[e.g.][]{komatsu:wmap5}.

\section{Secular Evolution versus Cosmological Evolution}
\label{sec:cosmo}

Consider an ``initial'' equilibrium, axisymmetric disk+halo system 
at time $t=0$. In this limit the system will not evolve any 
non-axisymmtric modes. Therefore, introduce a non-axisymmetric 
perturbation to the disk potential of 
amplitude
\begin{equation}
\delta_{0}\equiv \frac{\delta\phi}{\phi}\ .
\end{equation}
We are specifically interested in {\em global} models, so 
$\phi\sim G\,M/R$ is the potential of interest 
(where $M$ is the disk+enclosed halo mass and 
$R$ is a characteristic effective radius/scale length). 
The precise meaning of the perturbation $\delta\phi$ differs depending 
on the mode(s) of interest and configuration. For example, 
in idealized N-body simulations, this typically corresponds to 
shot noise. However, in realistic cosmological settings this will 
correspond to substructure in the disk or halo, with 
$\delta\phi\sim G\,m/r$ (where $m$ is the substructure 
mass and $r$ its ``initial'' distance). 
The relevant numerical prefactor will depend on the orbit, 
phase-space structure, and mode (for example, for a bar, 
the desired quantity is the time-averaged contribution to the 
$m=2$ mode at radius $\sim R$ in the co-rotating frame); for our purposes, 
the scaling (not absolute value) of $\delta$ is most important. 

At early times (before saturation), this non-axisymmetric term will be amplified 
internally and grow roughly exponentially:
\begin{equation}
\delta(t) = \delta_{0}\,\exp(t/t_{0})
\label{eqn:delta}
\end{equation}
where $t_{0}$ is the effective secular timescale, which is typical of order a few orbital 
times \citep[again, this is for global modes, not local; see e.g.][]{
holley:bar.halo.interaction,weinberg:bar.res.1,
weinberg:bar.res.requirements}. 
This growth time 
has been the focus of considerable study, 
and is one of the many important results of isolated 
disk studies. 
For example, for a disk bar in a strongly unstable bulge-free Milky Way-like 
disk, \citet{dubinski:bar.evol.sim.tests} show Equation~(\ref{eqn:delta}) 
is a good approximation 
to the behavior in simulations, with $t_{0}=8\pi/\kappa\approx2.83\,P_{d}$ 
(where $\kappa$ is the epicyclic frequency, $=2^{3/2}\pi\,P_{d}^{-1}$ 
for a constant circular velocity disk, and $P_{d}=2\pi\,R/V_{c}$ is the disk circular period 
at its effective radius). 
\citet{klypin:bar.dynamics.vs.thickness} 
find a similar $t_{0}\approx3-5\,P_{d}$ for 
thin, bulge-free MW-like disks \citep[albeit with a much larger 
$t_{0}\sim 10-30\,P_{d}$ for thick $H/R\gtrsim0.5$ disks; 
see also][]{colin:bar.in.mw.like.halo.evol}.
\citet{martinezvalpuesta:recurrent.buckling} 
see timescales from $\sim 2.5-10\,P_{d}$, depending on whether 
the bar growing is an initial mode or a secondary (post-buckling) mode. 
A similar range of timescales is found (with considerable 
galaxy-to-galaxy variation) in live cosmological halos 
in \citet{berentzen:disk.bar.in.live.halos}.

For less cosmologically motivated, but more general and 
analytically tractable disk mass profiles, 
\citet{athanassoula.sellwood:bar.timescale}
find typical $t_{0}\sim 1.0-6.7\,P_{d}$ for realistic halo mass 
fractions $\sim1/4-1/2$ (fraction of the total mass 
owing to the halo at $<R$) and scale heights $H/R\sim 0.1$.
\citet{narayan:87.bar.tscale.shearing.sheet} and 
\citet{shu:gas.disk.bar.tscale} obtain $t_{0}\sim 0.8-1\,P_{d}$ 
for gas disks with an outer Lindblad 
resonance at $R\gtrsim R_{e}$ (of interest for global modes here) 
and no halo. 

More stable systems will evolve more slowly; for the sake of 
generality we define
\begin{equation}
t_{0} = N_{\rm Disk\ Periods}\times P_{\rm d} \equiv \frac{1}{{1-\stabilityparam}}\,P_{\rm d}
\label{eqn:Xeff}
\end{equation}
where $\stabilityparam$ is an effective stability parameter: 
$\stabilityparam\sim 0$ represents typical, cosmologically realistic 
disks maximally unstable to large-scale modes, which will evolve on a 
single orbital time; and 
$\stabilityparam >1$ systems are stable and experience only oscillations, 
rather than amplifying modes.\footnote{
Under 
certain restrictive circumstances, our 
$\stabilityparam$ here is analogous to the Toomre Q or X parameter 
$X\equiv \kappa^{2}\,R/(2\,\pi\,n\,G\,\Sigma)$ or 
a (renormalized) Ostriker-Peebles criterion (proportional to 
the ratio of rotational kinetic to potential energy). 
For e.g.\ the bar in a two-dimensional Kuz'min disk 
approximation presented in \citet{athanassoula.sellwood:bar.timescale}, 
we can translate their Equation~3 to obtain 
\begin{equation}
\stabilityparam = 0.3 + 1.1\,{\Bigl(} \frac{f_{\rm halo}+f_{\rm bulge}}{1/3} - 1 {\Bigr)} 
+ 0.6\,{\Bigl(} \sqrt{\frac{H}{0.1\,R}} - 1 {\Bigr)}\ ,
\end{equation}
in physical terms of the disk thickness $H/R$ and halo plus bulge 
(non-disk) mass fraction inside $R$. 
The definition in Equation~\ref{eqn:Xeff} is not, however, meant to represent 
specific instabilities, but to allow for general large-scale disk modes 
with a characteristic growth time/stability criterion.
} 
Note that, formally speaking, $\stabilityparam<0$ is allowed. For certain 
bar configurations, for example, $t_{0}\sim0.7\,P_{d}$ 
has been obtained \citep[see e.g.][]{adams89:eccentric.instab.in.keplerian.disks,
earn.sellwood:95.nbody.bar.stab}, 
or even, for spiral structure in the weak 
winding approximation, $t_{0}\sim0.4\,P_{d}$ \citep{toomre:spiral.mode.growth}. 
However, those situations all involve no halo and an infinitely 
thin disk, and somewhat different matter profiles 
from what are observed in typical disks. 
For moderate halo contributions or disk thickness, 
$t_{0}$ is unlikely to be smaller than $P_{d}$ by any but a small 
factor ($t_{0}\sim 0.7-0.8\,P_{d}$), a small difference relative to the 
uncertainties in other quantities calculated here. 
And the MW-like examples above illustrate that $\stabilityparam\approx0.5-0.75$ 
is probably the case of greatest interest for 
realistic disk plus halo systems, even for strongly 
unstable systems. To be conservative, however, 
we will adopt $\stabilityparam=0$ for all numerical estimates, 
unless explicitly otherwise specified.

\begin{figure*}
    \centering
    \scaleup
    \plotside{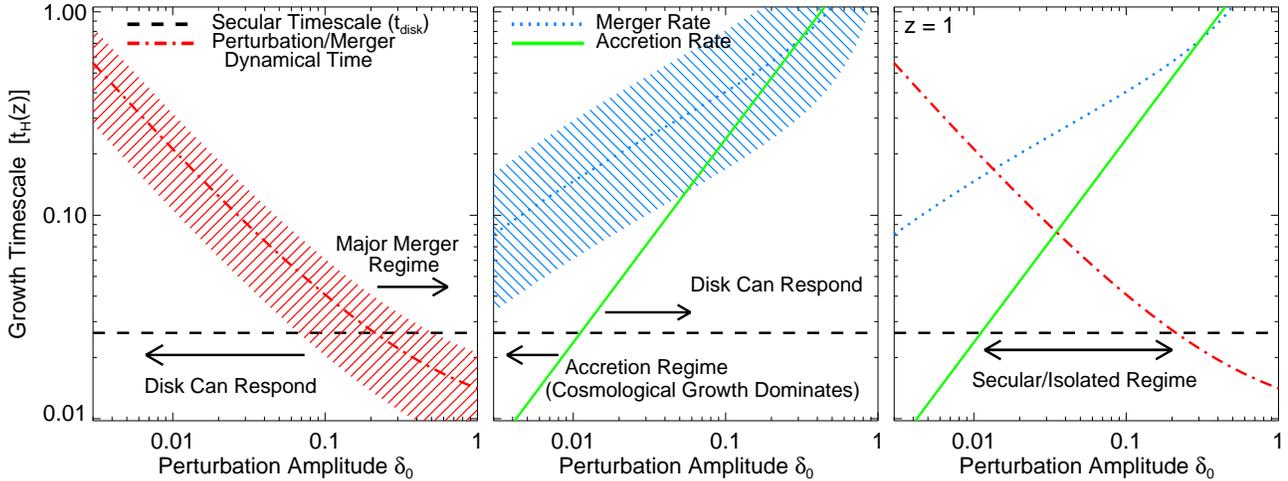}
    \caption{Characteristic timescales for 
    evolution of perturbations in unstable $\sim L_{\ast}$ 
    disks (here at $z=1$, with $\stabilityparam=0$). {\em Left:} Timescale for 
    internal disk response (secular evolution) to amplify some large-scale 
    mode with amplitude $\delta$, compared to the timescale for 
    an individual perturbation to evolve on its own (via e.g.\ dynamical friction). 
    Analogous to major mergers, direct evolution is more rapid than disk 
    response for major perturbations $\gtrsim0.2$. 
    Units are the age of the Universe at this redshift. 
    {\em Center:} Secular timescale versus timescale for the disk to accrete 
    a {\em new} fractional gas mass $>\delta$ or undergo a {\em new} merger 
    with mass ratio $>\delta$. At sufficiently low amplitudes, accretion is 
    non-negligible over the secular response timescale. 
    {\em Right:} All timescales. Disks are effectively 
    {\em both} isolated and potentially secular evolution-dominated in a 
    regime around $\delta\sim0.1$. Raising $\stabilityparam$ will increase the 
    ``secular timescale'' and decrease this range. 
    \label{fig:compare}}
\end{figure*}

However, galaxies are not static, and two things will happen that might compete with 
this internal self-amplification: (1) the substructure itself can dynamically evolve, 
driving stronger 
perturbations and/or merging; and (2) 
new mass of magnitude comparable to the disk mode can be 
accreted/merge. If either of these occurs on a timescale shorter than $t_{0}$ 
(the effective secular timescale), the system should not be considered 
``isolated'' for purposes of secular evolution. 

Consider case (1), the dynamical evolution of the substructure itself. 
Given some substructure/perturbation of mass fraction $\delta$ 
at some initial radius of 
interest $r$, the orbit will decay on a timescale of order the dynamical friction 
time; correspondingly, the perturbation $\delta\propto\delta\phi\propto r^{-1}$ 
will grow on the same timescale.\footnote{
Strictly speaking, 
realistic cosmological perturbations grow continuously, so an 
``initial'' radius is ambiguous. However, there is still some 
$\delta\phi$ that scales as described at a given instantaneous 
$r$, and this is what ultimately enters into the equations derived. 
Also, in practice, such modes -- where induced by substructure -- 
often appear suddenly (i.e.\ in a 
time $<P_{d}$ when $r\sim R$; this is because at larger radii, 
the {\em net} non-axisymmetric $\delta\phi$ contribution is suppressed 
by a Poisson $\sim N^{-1/2}$ ($\sim R^{-3/2}$) term. In simulations, for example, 
perturbations are typically dominated by a few close passages of 
clumps/substructure where $r\sim R$ \citep[although these may be from 
longer radial orbits; see][]{velazquezwhite:disk.heating,
bournaud:gas.bar.renewal.destruction,
gauthier:triggered.bar.from.substructure,
kazantzidis:mw.merger.hist.sim,hopkins:disk.heating}. 
In any case, since our derivations rely on 
$\delta$, rather than $r$ explicitly, this is not a large source of uncertainty.
}
Strictly speaking, dynamical friction
does not dominate angular momentum loss at small radii; rather, 
resonant tidal interactions act more efficiently \citep{barneshernquist92}. 
However, properly 
calibrated, the dynamical friction time is not a bad approximation 
\citep[e.g.][]{boylankolchin:merger.time.calibration}. For an 
isothermal sphere or \citet{mestel:disk.profile} (flat rotation curve) disk, this time 
is simply: 
\begin{equation}
t_{\rm df} = \frac{R/V_{c}}{2\,\beta\,\ln{\Lambda}}\,\frac{M_{\rm enc}(r)}{m}\,\frac{r}{R}
\approx \frac{0.2}{\delta_{0}\,\ln{\Lambda}}\,P_{\rm d} 
\label{eqn:df}
\end{equation}
where the equality on the right comes from the definitions 
of $\delta_{0}$ and $P_{d}$.\footnote{
In detail, 
$\beta$ is a constant that weakly depends on the mass profile 
and velocity isotropy: $=0.428$ for an isotropic isothermal 
sphere and $=0.32$ for a thin \citet{mestel:disk.profile} disk averaged 
over random inclinations (used in 
Equation~\ref{eqn:df}). The Coulomb logarithm is 
approximately $\Lambda=1+1/\delta_{0}$ \citep{boylankolchin:merger.time.calibration,
jiang:dynfric.calibration}. 
For Figure~\ref{fig:compare}, we use the fitting functions from 
\citet{boylankolchin:merger.time.calibration}, with appropriate eccentricity and 
orbital parameter dependence, rather than the simplified 
Equation~\ref{eqn:df}, but the results are similar on average. 
}
Since we are considering the magnitude of the perturbation relative to the 
disk, the time here scales with the disk dynamical time at fixed 
$\delta_{0}$ (as opposed to e.g.\ the Hubble time 
for halo-halo orbital decay at large radii). 

The left panel of Figure~\ref{fig:compare} compares this timescale
to the secular evolution timescale $t_0$. For representative purposes, 
we assume a ``maximally unstable'' $t_{0}=P_{d}$ ($\stabilityparam=0$) 
MW-like disk with $P_{d}=2\,\pi\times5\,{\rm kpc}/200\,{\rm km\,s^{-1}}\approx 160$\,Myr, 
and total stellar mass $=5\times10^{10}\,\msun$. 
This is easily generalized; $P_{d}$ (at the 
scale $\sim R$ of the disk itself) appears to be independent of 
mass in observed disks \citep[e.g.][and references therein]{courteau:disk.scalings}. 
We plot the results assuming such a disk exists at redshift $z=1$, 
but the qualitative scalings are similar at redshift $z=0$, 
and we will show the redshift dependence explicitly below. 
We compare the dynamical friction time $t_{\rm df}$; 
here we show the results using the full orbital parameter-dependent 
fits from simulations in \citet{boylankolchin:merger.time.calibration}, 
which allows us to quote 
the $\pm1\,\sigma$ range of $t_{\rm df}$ from the range of 
orbits observed in cosmological simulations \citep{benson:cosmo.orbits,
khochfar:cosmo.orbits}. Using the simpler formula in Equation~\ref{eqn:df} 
is similar to the median expected. 

Comparing Figure~\ref{fig:compare} or Equations~\ref{eqn:Xeff} \&\ \ref{eqn:df}
shows that the dynamical evolution of the perturbation is more 
rapid than the internal response for mass ratios larger than 
\begin{equation}
\delta_{\rm crit,\, df} = \frac{{1-\stabilityparam}}{4\,\pi\,\beta\,\ln{\Lambda}} 
\sim 0.2\,(1-\stabilityparam)\  . 
\end{equation}
This is ultimately an obvious regime; when $\delta\phi/\phi\sim1$, direct evolution 
dominates the potential fluctuations. We denote this the ``major merger regime'': 
in the case where $\delta$ corresponds to some substructure, 
this clearly requires a mass ratio $\mu\gtrsim0.2$ with $r\sim R_{d}$, i.e.\ 
close passages of major companions. Note though that this does not 
{\em have} to be a merger. For example, a sufficiently strong disk 
fragmentation event will be similar. Physically, this is still dynamically distinct 
from secular evolution (from e.g.\ bars, etc.) -- it will ``look like'' a merger 
inside the disk \citep[see e.g.][]{elmegreen:classical.bulges.from.clumps}. 

Now consider case (2): new growth/perturbations/mergers. 
Note that we are no longer considering the evolution of 
{\em individual} perturbations, but the time between new perturbations of 
the same or greater magnitude. If this is $\ll t_{0}$, then the system 
is not isolated. A lower limit to this is given by the rate of 
{\em baryonic} accretion/merging onto the disk (if accreted systems 
retain some dark matter, they will represent larger perturbations, 
but there is at least a lower-limit in the mass added in baryons to explain the 
disk mass). Detailed analyses of these rates have been 
discussed extensively in the literature \citep[see e.g.][]{brown:mf.evolution,
guo:gal.growth.channels,wetzel:mgr.rate.subhalos,
genel:smg.numden.vs.mergers,stewart:mw.minor.accretion}.
Here, we use a simple semi-empirical model to define some 
of the relevant scalings; for more discussion, see \citet{hopkins:disk.survival.cosmo}. 
A variant of the model, based on subhalo-subhalo merger rates, is also 
described in detail in \citet{hopkins:groups.ell}.
Following \citet{stewart:merger.rates}, we begin 
with dark matter halo merger trees \citep[here from][]{fakhouri:halo.merger.rates}. 
Empirical halo occupation models 
and other observations constrain the average galaxy mass per host halo (or 
subhalo) mass, with little scatter -- so at a 
given instant we simply populate the halos with galaxies.  
Here assigning stellar mass given the fitted 
$M_{\ast}(M_{\rm halo}\,|\,z)$ from 
\citet{conroy:hod.vs.z} and gas mass given the fits 
to $M_{\rm gas}(M_{\ast}\,|\,z)$ from 
\citet{stewart:disk.survival.vs.mergerrates} \citep[for the observations used in the 
fits, see references therein and][]{belldejong:tf,erb:lbg.gasmasses,
fontana:highz.mfs,perezgonzalez:mf.compilation}. 

The uncertainties in this modeling methodology 
will be discussed in detail in \citet{hopkins:merger.rates}, but for 
our purposes they are relatively small 
(factor $\sim2$ uncertainty in the merger rate near $\sim L_{\ast}$, 
owing to a combination of uncertainty in $M_{\ast}(M_{\rm halo})$ 
and the halo-halo merger rate) at $z<2$, because 
it is primarily the {\em shape} of the galaxy-halo mass correlation 
(rather than e.g.\ its absolute normalization) that affects 
galaxy-galaxy merger rates.\footnote{The merger rates 
from this model as used here can be also 
obtained as a function of e.g.\ galaxy 
mass, mass ratio, and redshift from the ``merger rate calculator'' 
script publicly available at 
\mergercalcurl.} Note, however, that the 
uncertainties grow rapidly at higher redshifts, owing to the lack of 
empirical constraints. Evolving the system 
forward some small increment in time, we can ``add up'' the mergers 
\citep[in detail, we add a dynamical friction ``delay'' time between each 
halo-halo merger and subsequent galaxy-galaxy merger, with the 
formulae from][]{boylankolchin:merger.time.calibration}. 
This gives merger rates; but also, knowing the new halo mass (after 
accretion/growth in this time interval), 
the empirical halo occupation constraints define the ``expected'' 
galaxy mass for the updated halo mass. We simply assign whatever 
galaxy mass growth is needed to match this (not already brought in 
by mergers) to ``accretion.'' Note that this is a lower limit to the 
accretion rate, reflecting {\em net} accretion (outflows may remove mass, 
requiring more new gas inflow). 

The middle panel of Figure~\ref{fig:compare} shows the relevant timescale for both mergers 
(median time $\Delta t$ between mergers with baryonic mass ratio 
$\mu\equiv M_{\rm bar,\,2}/M_{\rm bar,\,1} > \delta_{0}$) 
and accretion 
($\Delta t$ for the disk to grow via accretion by a mass fraction $>\delta_{0}$). 
Accretion tends to be the dominant growth channel (relative to e.g.\ minor 
mergers), for all but the most massive galaxies (where gas accretion 
is ``quenched''). As a result, the time between new mergers may be long, but 
at sufficiently low $\delta_{0}$, growth by accretion is more rapid than 
internal disk response. 
We denote this the ``accretion regime.'' Again, the behavior is 
easily understood: if one is interested in 
evolution at the $\ll 10\%$ level, then 
galaxies cannot be considered isolated 
for even a single dynamical time, as they will grow 
by more than this amount in that time. 

The relevant criterion can be roughly estimated as follows: to very crude approximation, 
fractional galaxy growth rates scale as $\sim \alpha/t_{\rm Hubble}$, 
where $\alpha$ is weakly redshift dependent but non-trivially 
mass-dependent with $\alpha\sim0.2$ for 
a Milky-Way mass halo at $z=1$ (i.e.\ an 
assumed galaxy mass $5\times10^{10}\,\msun$). 
For such a system, as pictured in Figure~\ref{fig:compare}, the galaxy will grow by a fraction 
$>\delta_{0}$ in the time $t_{0}$ (secular response time) 
for perturbation amplitudes below 
\begin{equation}
\delta_{\rm crit,\ acc} = \frac{\alpha}{{(1-\stabilityparam)}}\,\frac{P_{d}}{t_{H}(z)} \sim 
\frac{0.003}{{1-\stabilityparam}}\ (z=0)\ .
\label{eqn:muz.cosmo.z0}
\end{equation}

\begin{figure}
    \centering
    \scaleup
    \plotone{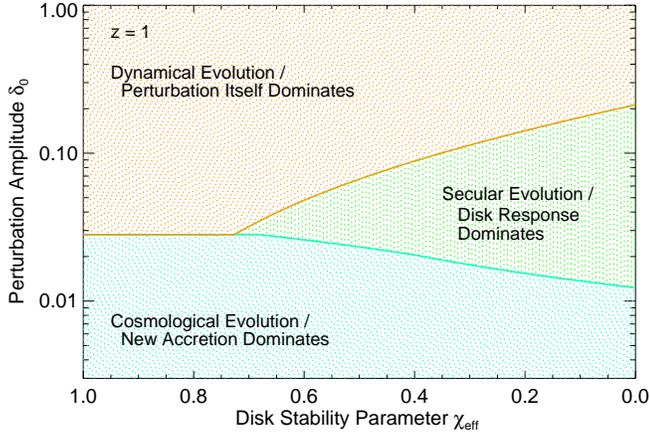}
    \caption{Parameter space of regimes in Figure~\ref{fig:compare} for 
    the same $\sim L_{\ast}$ system as a function of perturbation amplitude and effective 
    disk stability (speed of the growth of secular modes of interest). 
    \label{fig:param.space}}
\end{figure}

Figure~\ref{fig:compare} considers the ``maximally unstable''  
($\stabilityparam=0$) case, such that $t_{0}=P_{d}$. If the stability parameter 
is higher (larger $t_{0}$), the regime of effective ``isolation'' will 
be more restricted. 
Figure~\ref{fig:param.space} illustrates the parameter 
space as a function of 
the effective disk stability parameter $\stabilityparam$ 
(recall, this is simply defined relative to the number or orbits 
needed to grow the mode of interest). Above some critical 
$\stabilityparam$ (here $\stabilityparam\sim0.75$, i.e.\ $N_{\rm orbits}=4$ or 
$t_{0}\gtrsim0.5$\,Gyr for a MW-like disk), the secular timescale is 
always longer than the other timescales above. This is simply the 
statement that disks are not ``isolated'' for 
timescales $\gtrsim$\,Gyr, especially at high redshift.

\section{Dependence on Galaxy Mass and Redshift}
\label{sec:mz}

\begin{figure*}
    \centering
    \plotside{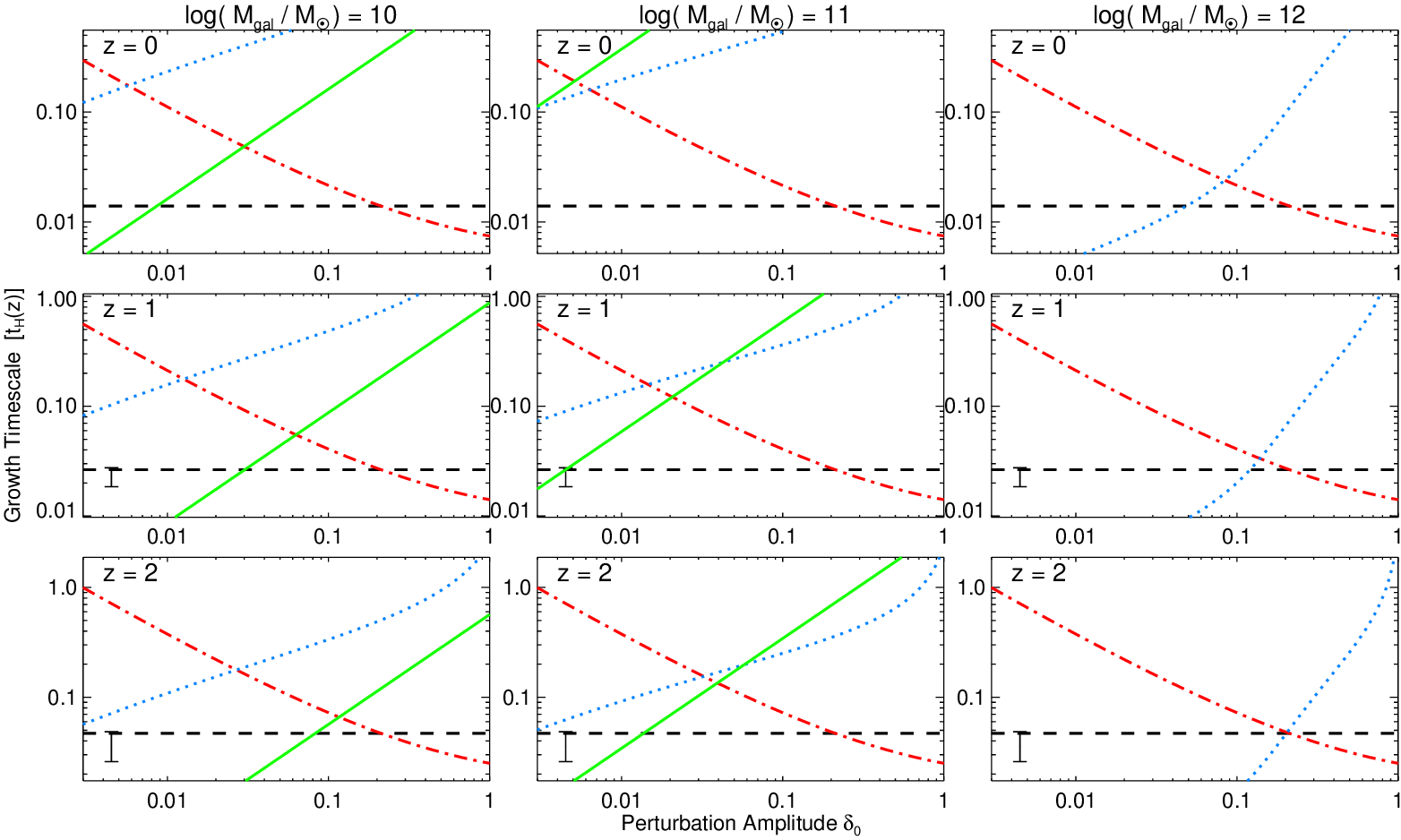}
    \caption{Same as the right panel of Figure~\ref{fig:compare} ($\stabilityparam=0$), 
     as a function of 
    galaxy mass and redshift. 
    In low-mass ($\lesssim10^{10}\,\msun$; {\em left}) galaxies, 
    merger rates are low, but accretion rates rapid -- secular responses 
    at the $\ll 10\%$ level compete with cosmological disk growth. 
    At intermediate masses $\sim L_{\ast}$ ($\sim10^{11}\,\msun$; {\em center}) 
    accretion and minor mergers occur with comparable rates. 
    At high masses ($\gtrsim 10^{11}\,\msun$; {\em right}) 
    accretion rates are low (cooling is inefficient) but merger rates 
    grow rapidly -- secular responses at the $\lesssim 10\%$ level 
    compete with mergers. 
    As a function of redshift, disk dynamical times scale weakly, 
    but merger and accretion rates increase, leaving less 
    of parameter space in which disks can be considered ``isolated'' for 
    the internal response time. Error bars mark the range 
    between the internal response time if disk sizes do not evolve 
    with redshift ($\beta_{d}=0$; dashed lines) and if they 
    evolve at the maximum rate constrained by 
    observations ($\beta_{d}=0.6$; lower bar). 
    \label{fig:compare.m.z}}
\end{figure*}

Figure~\ref{fig:compare.m.z} shows how the regime of 
secular evolution depends on galaxy mass 
and redshift. First, we consider the 
same comparison at $z=0$ as a function of galaxy mass. 
Observations indicate that $P_{d}$ is nearly mass-independent 
at the disk effective radii of interest for global models \citep{belldejong:tf,shen:size.mass,
courteau:disk.scalings}. Given Equation~\ref{eqn:df}, the same is true for 
dynamical evolution of individual perturbations (at fixed $\delta_{0}$). However, 
accretion and merger rates scale significantly with mass. At low masses, 
merger rates are low, but accretion rates are high. At high masses, accretion 
rates drop rapidly (consistent with zero at $M_{\rm gal}\gg 10^{11}\,\msun$), 
but merger rates increase, leaving almost no range of perturbation in which
secular processes are relevant (right column of Figure~\ref{fig:compare.m.z}).  
Both effects are seen in a variety of 
models and observations \citep{maller:sph.merger.rates,
noeske:sfh,guo:gal.growth.channels,kitzbichler:mgr.rate.pair.calibration,
parry:sam.merger.vs.morph,
stewart:merger.rates,keres:cooling.revised,bundy:merger.fraction.new}. 
The mass dependence is important even over a relatively narrow mass 
range -- for example, note that our 
previously assumed Milky-Way like mass of $5\times10^{10}\,\msun$ (Figure~\ref{fig:compare}), 
being a factor $\sim2$ smaller than the $10^{11}\,\msun$ case shown here, 
has correspondingly more rapid accretion rates (between the 
$10^{10}\,\msun$ curve and $10^{11}\,\msun$ curve). 

Again, we emphasize that we are using baryonic mass ratio 
$\mu$ here -- this is a minimum, 
as it reflects the most densely bound material that will survive to 
perturb the galaxy \citep[an individual merger may ``begin'' at larger $\delta$ including 
dark matter, or smaller $\delta$ at large radii, but orbital decay and 
stripping will tend to saturate it at $\delta\phi/\phi\sim \mu$, with a rate of 
new such events from mergers as shown; see e.g.][]{kazantzidis:mw.merger.hist.sim}. 
Low-mass galaxies are observed to be more dark-matter dominated, so 
if this can be conserved, the relevant rates will not decrease as rapidly with stellar 
mass; however, modeling this requires more detailed knowledge of 
cosmological orbits, stripping, and internal galaxy structure. 

For each mass, Figure~\ref{fig:compare.m.z} shows how the regime of 
secular evolution depends on redshift. 
To lowest order, accretion timescales 
evolve with the Hubble time 
\citep[fitting directly, accretion rates 
$\propto (1+z)^{2}$; see][]{stewart:merger.rates}. 
Observations of the 
baryonic Tully-Fisher and size mass relation suggest that 
$P_{d}$ (or equivalently at fixed mass, disk sizes) 
evolves weakly from 
$z=0-2$ \citep[][]{trujillo:size.evolution,flores:tf.evolution,
kassin:tf.evolution,toft:z2.sizes.vs.sfr,
akiyama:lbg.weak.size.evol,somerville:disk.size.evol}. 
Moreover, theoretical models that include the well-established 
dependence of halo concentration on redshift 
\citep[see e.g.][]{bullock:concentrations,
wechsler:concentration} predict a similar 
weak scaling \citep{somerville:disk.size.evol}. 
Parameterizing as $P_{d}\propto (1+z)^{-\beta_{d}}$, these observations 
constrain $\beta_{d}= 0.0 - 0.6$. In Figure~\ref{fig:compare.m.z}, we 
conservatively adopt $\beta_{d}=0$ (i.e.\ $P_{d}$ independent of redshift), 
but we show how the results would change if we allowed the 
maximum observationally inferred evolution, $\beta_{d}=0.6$. 
It makes a small difference, but does cancel some of the redshift evolution 
in the relevant parameter space. Even in the extreme case of 
a simple $P_{d}\propto t_{\rm Hubble}$ scaling \citep{momauwhite:disks}, 
some, but not all of the evolution is negated (at $z<2$, 
merger and cooling rates evolve as $\propto(1+z)^{2}$, 
$1/t_{\rm Hubble}$ as $\propto (1+z)$). 

In Figure~\ref{fig:compare.m.z}, the critical amplitude 
below which the ``accretion regime'' pertains  
scales {\em roughly} as $\delta_{\rm crit,\ acc}\propto (1+z)^{1.5-2.0}$, 
while $\delta_{\rm crit,\, df}\sim$\,constant. 
This is an approximation over the entire range $z=0-2$; 
in fact at the lowest redshifts ($z\lesssim0.2$), the 
falloff in $\delta_{\rm crit,\ acc}$ is somewhat more rapid 
(as e.g.\ the Universe's acceleration term becomes important). 
As a consequence, the 
range of $\delta_{0}$ over which ``isolation'' is a good 
approximation decreases with increasing redshift.

\begin{figure}
    \centering
    \scaleup
    \plotone{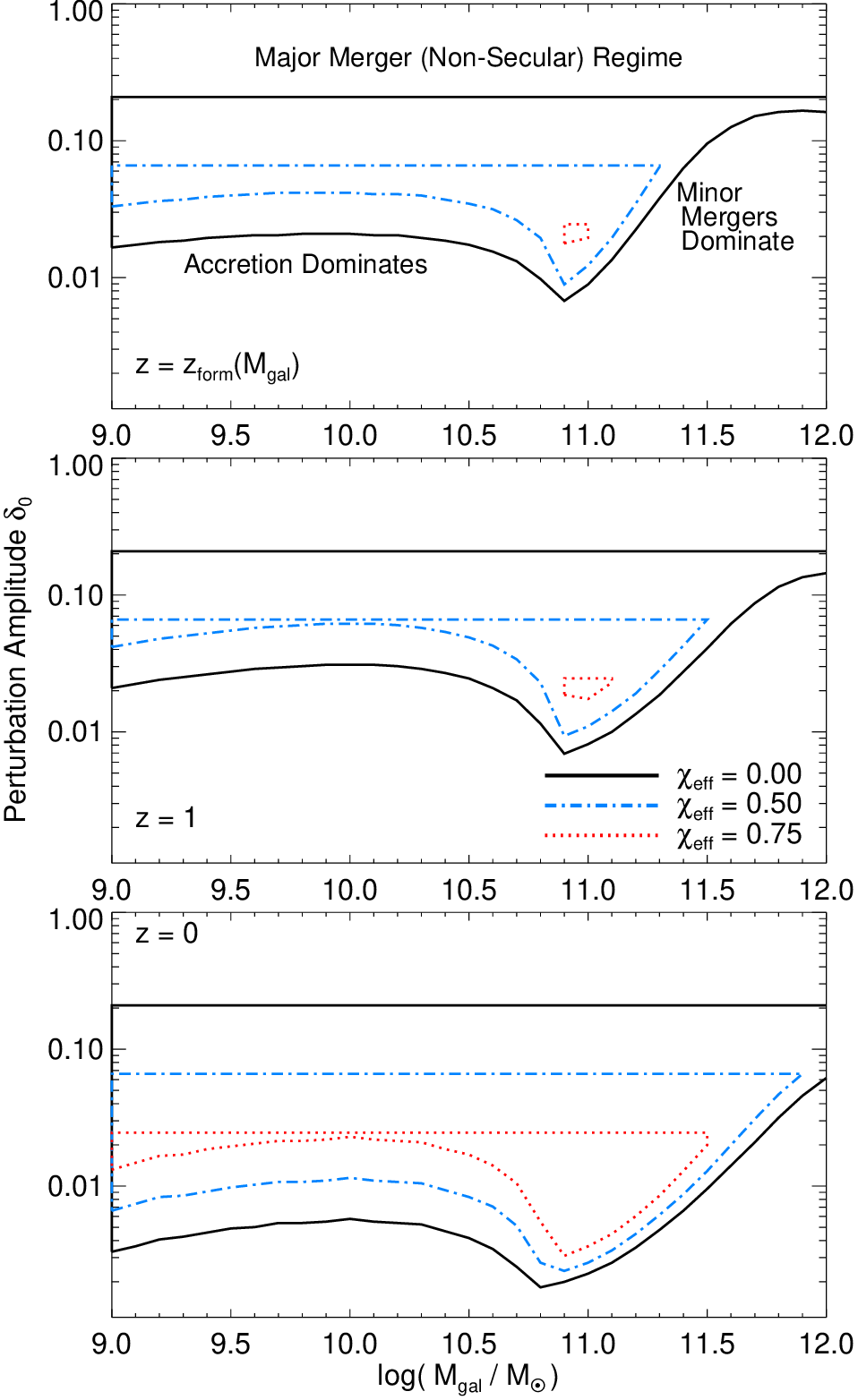}
    \caption{Parameter space of the ``isolated'' secular regime 
    (Figure~\ref{fig:compare}) versus galaxy mass and $\stabilityparam$, 
    for a redshift-independent disk period $P_{d}$.
    {\em Top:} At the formation redshift $z=z_{\rm form}$, 
    where each galaxy reaches half its $z=0$ mass. 
    {\em Middle:} At $z=1$. 
    {\em Bottom:} At $z=0$.
    \label{fig:param.space.avail}}
\end{figure}

Figure~\ref{fig:param.space.avail} summarizes the parameter space 
as a function of galaxy mass and stability parameter $\stabilityparam$, 
at $z=0$, $z=1$, and $z=z_{\rm form}(M_{\rm gal})$. 
We define $z_{\rm form}(M_{\rm gal})$, the galaxy assembly time, 
as the redshift when each galaxy reaches half its $z=0$ mass, 
according to our simple growth model. 
To the extent that secular modes are considered important in this 
formation process, this is an interesting timescale.

\begin{figure}
    \centering
    \scaleup
    \plotone{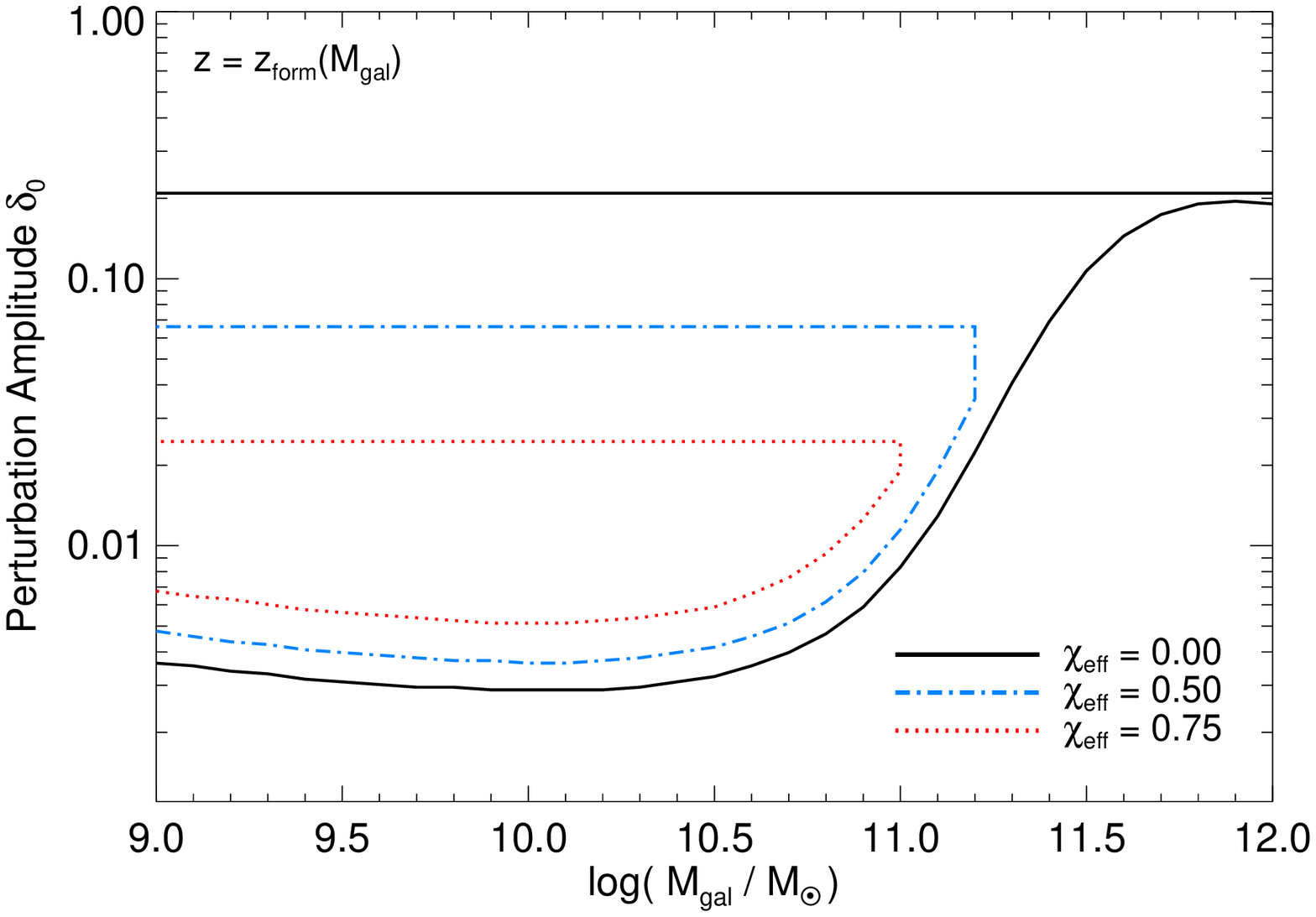}
    \caption{As Figure~\ref{fig:param.space.avail} (with $z=z_{\rm form}$); 
    but neglecting cosmological accretion (i.e.\ considering only merger 
    timescales as the competing timescale at low-$\delta$). 
    \label{fig:param.space.noacc}}
\end{figure}

Simulations find that star-forming galaxies accrete most of their mass 
along a couple of dynamically coherent, clumpy filaments; as such 
they are dynamically important for large-scale disk modes 
\citep{keres:hot.halos,keres:cooling.revised,dekelbirnboim:mquench,
dekel:cold.streams}. If, however, accretion were perfectly 
smooth, axisymmetric, and restricted to large radii (without 
migration of new material inwards), then it might be valid to ignore 
it in studying secular modes even when accretion rates are large. 
To represent this possibility, Figure~\ref{fig:param.space.noacc} 
re-calculates Figure~\ref{fig:param.space.avail}, but ignores accretion. 
At low masses, merger rates are sufficiently low that the isolated regime 
extends to smaller mass ratios $\delta<0.01$.

\section{Implications for Mode ``Lifetimes''}
\label{sec:lifetimes}

\begin{figure}
    \centering
    \scaleup
    \plotone{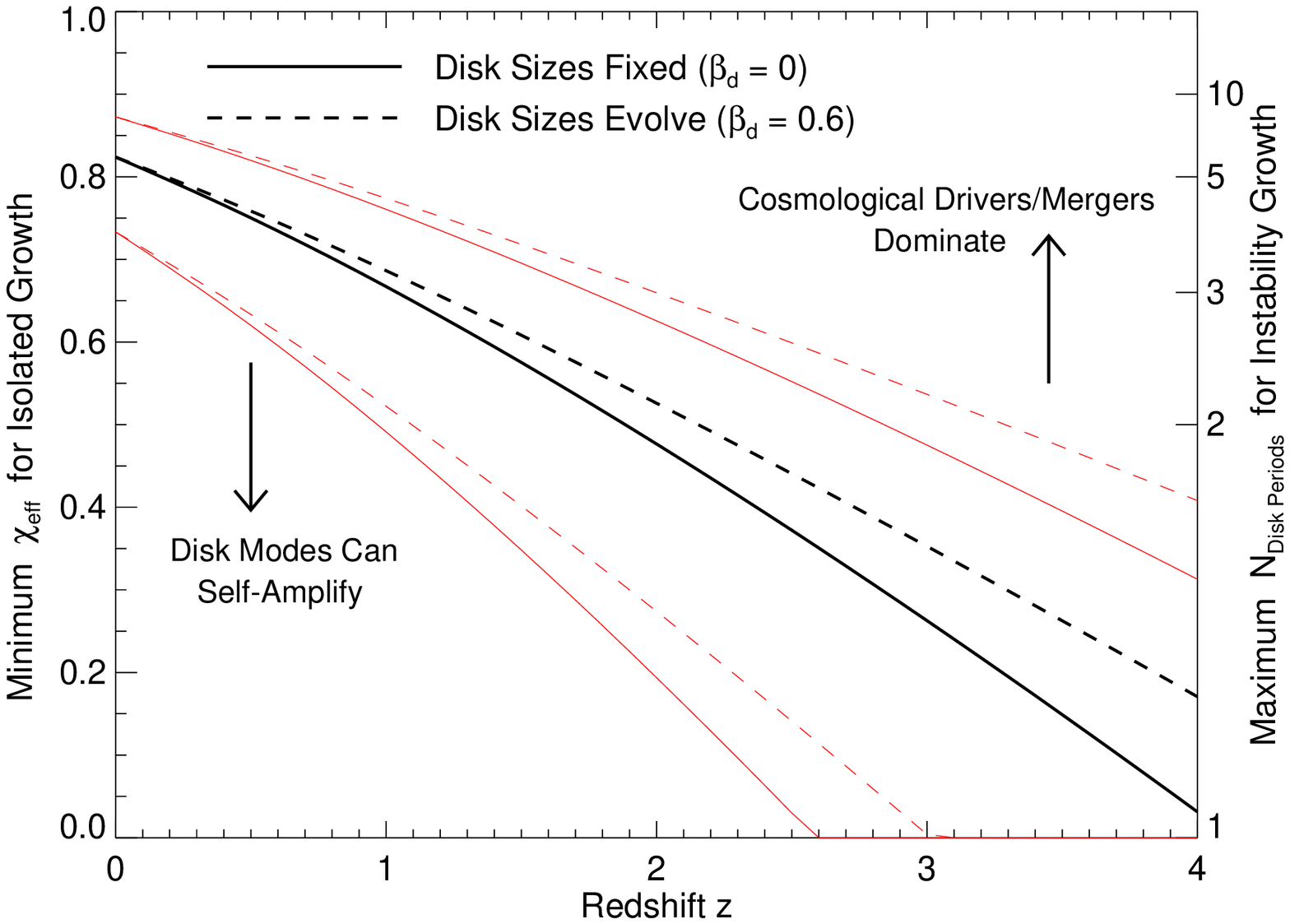}
    \caption{Minimum effective stability parameter $\stabilityparam$ or 
    maximum $N_{\rm Disk\ Periods}\equiv 1/(1-\stabilityparam)$ where 
    the secular growth timescale equals the maximum isolated lifetime 
    in Figure~\ref{fig:bar.lifetime.limit}. Black (red) lines show the 
    median ($16-84\%$ range) expected. In more stable (slower-responding) 
    disks, large-scale modes should be considered cosmologically 
    dynamical systems. 
    \label{fig:X.limit}}
\end{figure}

The cosmological evolution of galaxies 
also has important implications for mode ``lifetimes.'' 
Since $t_{0}=1/(1-\stabilityparam)\,P_{d}$, there is clearly some $\stabilityparam$ 
at each redshift above which $t_{0}$ is larger than any of the competing 
timescales for {\em all} $\delta_{0}$. Modes with 
larger $\stabilityparam$ are 
still formally unstable, but the time/number of orbits to amplify the mode 
becomes sufficiently long that these modes should be 
considered cosmologically dynamical objects. 
Figure~\ref{fig:X.limit} shows this maximum $\stabilityparam$ as a function 
of redshift (for $\sim10^{11}\,\msun$ galaxies where this 
is maximized, as seen in Figure~\ref{fig:param.space.avail}). 
At $z\gtrsim1$, this corresponds to modes 
growing in $\lesssim$ a couple $P_{d}$; at $z\ge2$, however, even 
$\stabilityparam=\ll 1$ systems (those where modes grow on a timescale 
$\sim P_{d}$) can be in the ``accretion regime,'' as discussed above. 
Recall, simulations suggest that even 
cold, bulge-free MW-like disks have effective $\stabilityparam\sim0.5-0.75$ 
\citep[][and references therein]{dubinski:bar.evol.sim.tests}. 
This high-$z$ behavior is directly related to observations showing that disk orbital 
periods at high redshifts become comparable to the 
Hubble time \citep[see e.g.][]{flores:tf.evolution,
kassin:tf.evolution,toft:z2.sizes.vs.sfr,vanstarkenburg:z2.disk.dynamics,
shapiro:highz.kinematics}.

At $\stabilityparam$ less than the values above, secular modes can grow 
``in isolation'' from some $\delta_{0}$. 
Typically, these will grow rapidly 
and saturate at some $\delta_{f}\sim1$. 
However, if an isolated mode then survives stably 
at an amplitude $\delta_{f}$ for a lifetime much longer than the 
other timescales compared here, then various cosmological effects 
may have important consequences. For example, if a disk bar 
saturates and survives with some $\delta_{f}\sim0.4$ \citep{dubinski:bar.evol.sim.tests}, 
in some number of dynamical times the galaxy will grow by this much. 
Essentially, cosmological growth may ``catch up'' to the saturated 
mode and could effect it. Of course, the mode could continue growing with the 
galaxy, or be robust to these effects; our point is that 
continuing to treat such a mode 
in isolation may not necessarily be a good approximation over 
much longer timescales. Moreover, if stable modes can survive for a 
timescale much longer than e.g.\ the relevant dynamical friction times at 
$\delta_{0}$, then the {\em presence} of those modes mode (the duty cycle) 
will de-couple from that of their drivers (if they were initially driven). In e.g.\ the 
case of minor mergers, this is the statement that new mergers 
and/or the destruction of the original driving satellites will wipe out 
the ``memory'' of the drivers, while the bar survives.

\begin{figure}
    \centering
    \scaleup
    \plotone{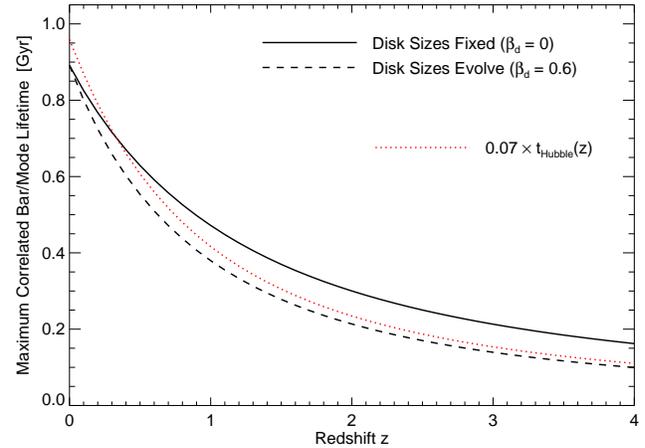}
    \caption{The maximum ``isolated'' lifetime of large-scale 
    modes (e.g.\ disk bars). This is the largest timescale (marginalizing over 
    $\delta$ and $M_{\rm gal}$) shorter than the competing (non-secular) 
    timescales in e.g.\ Figure~\ref{fig:compare.m.z}. Variations within populations 
    contribute factor $\sim2$ scatter from object-to-object. We show a 
    fixed fraction of the Hubble time for comparison. 
    Evolution on longer timescales 
    will compete with cosmological effects. 
    The duty cycle will de-couple 
    from driving: even if all such modes were driven by e.g.\ minor mergers, 
    there will be no correlation between the presence of the modes and the 
    presence of companions. 
    \label{fig:bar.lifetime.limit}}
\end{figure}

Taking the minimum of the non-secular timescales of interest 
(e.g.\ accretion and merger 
timescales in Figure~\ref{fig:compare.m.z}) at whatever amplitude $\delta$ 
maximizes this timescale, gives the maximum relevant 
``isolated'' mode lifetime. This is clearly a function of mass; we consider 
here the $\sim10^{11}\,\msun$ ($\sim L_{\ast}$) case, of greatest interest 
both as a MW-like system and because Figure~\ref{fig:param.space.avail} 
demonstrates that this is where such a timescale (the ``isolated'' regime) 
is maximized. Figure~\ref{fig:bar.lifetime.limit} plots this 
timescale versus redshift. We show this both for the assumption that 
$P_{d}$ does not evolve ($\beta_{d}=0$) and 
the maximum observationally constrained evolution ($\beta_{d}=0.6$). 
We compare a constant fraction ($\sim0.1$) of the Hubble time -- this appears 
to be a good approximation, on average (there will of course be 
scatter galaxy-to-galaxy in accretion and merger rates, leading to 
typical factor $\sim2$ scatter in the relevant timescale here).

\section{Discussion}
\label{sec:discussion}

Under typical cosmological conditions, global ``secular'' evolution -- narrowly 
defined as evolution by internal amplification of large-scale disk modes 
in effectively {\em isolated} galaxies -- only occurs in a restricted range of 
parameter space (Figures~\ref{fig:compare}-\ref{fig:param.space}). 
If the perturbation mode of interest has a fractional amplitude $\ll0.1$, what we 
call the ``accretion regime,'' then the disk will grow by accretion by a 
comparable amount in even a single dynamical time; the isolated 
approximation is clearly not valid.
This threshold is around an amplitude 
$\delta_{\rm crit,\ acc}\sim0.002\,(1-\chi_{\rm eff})^{-1}\,(1+z)^{1.5-2}$ for $10^{11}\,\msun$ 
galaxies (slightly lower at $z<0.2$) 
or $\delta_{\rm crit,\ acc}\sim0.005\,(1-\chi_{\rm eff})^{-1}\,(1+z)^{1.5-2}$ 
for $10^{10}\,\msun$ systems. 
At the opposite extreme, seed ``perturbations'' of fractional amplitude 
$\delta_{0} > \delta_{\rm crit,\ df}\sim0.2$ lead to non-secular evolution -- the 
perturbations' own gravitational evolution will dominate the internal 
response (this is obvious in the case of e.g.\ galaxy-galaxy major mergers 
or massive disk fragmentation events, where the evolution of the 
merger/clumps drives violent relaxation). 

The relevant parameter space depends on galaxy mass 
(Figures~\ref{fig:compare.m.z}-\ref{fig:param.space.noacc}). Although 
halo growth is nearly mass-independent 
\citep{fakhouri:halo.merger.rates,guo:gal.growth.channels,stewart:merger.rates}, 
galaxy growth histories 
are not (the function $M_{\rm gal}(M_{\rm halo})$ is non-trivial). 
At high masses ($M_{\rm gal}\gtrsim10^{11}\,\msun$) galaxy-galaxy merger 
rates are high such that systems are rarely ``isolated'' over the 
timescales of interest for secular evolution. However, there are few 
disks at these masses, so secular evolution is not expected to be 
a dominant process. At low masses ($\lesssim10^{10}\,\msun$) 
merger rates are low (in terms of galaxy-galaxy baryonic mass ratios; 
including dark matter, they may remain high) 
but accretion rates are high; systems can be 
effectively approximated as isolated for only a couple of orbits 
in the regime of amplitudes $\delta\sim0.03-0.2$. Moreover, 
although such galaxies are mostly disk ($B/T\ll 1$), 
they are increasingly dark matter-dominated 
which helps stabilize them 
to the development of secular modes \citep[see e.g.][]{persic96,
mihos:lsb.gal.stable.vs.bars,
borriello01,belldejong:tf}. 
Galaxies may be ``most isolated,'' and so traditional secular 
evolution most relevant, between these regimes, i.e.\ in galaxies somewhat 
below $\sim L_{\ast}$. That this occurs at masses only somewhat 
below where mergers become efficient is also interesting; there may be a 
relatively rapid regime (as galaxies approach and cross $\sim L_{\ast}$ 
in mass) in which today's galaxies transition from accretion-dominated, 
secularly stable (dark matter-dominated) disks, to 
secularly unstable (self-gravitating) disks, which could quickly 
amplify $\sim10\%$ amplitude perturbations into very strong 
bars and build significant pseudobulges, until later mergers destroy the remains of 
the disk and build massive classical bulges. 

This has important implications for the lifetimes of secular processes of 
interest. The above comparisons assume disks where the internal response 
occurs over a single orbital period; if the systems have higher effective 
stability (i.e.\ secular responses build more slowly), then the regime where 
they can be considered isolated for this time shrinks. Large-scale 
modes that require more than a few disk periods to self-amplify at low-redshift, 
or more than just a single disk period at high redshift ($z>2$), 
should be considered cosmologically dynamical systems (Figure~\ref{fig:X.limit}) -- 
the galaxy grows comparably over this self-amplification timescale. 
Indeed, various observations of 
disk sizes and structure suggest that disks are sufficiently thick or 
have sufficient bulge fractions such that internal response times 
are in this interesting range 
\citep{belldejong:tf,mcgaugh:tf,courteau:disk.scalings,
gilmore02:last.mw.merger.from.thick.disk,
wyse06:thick.disk.stars,barteldrees:exp.vertical.profiles,degrijs:vertical.disk.profiles}. 

Even if modes can evolve/self-amplify quickly 
such that a bar will grow efficiently and saturate at some final 
amplitude, these competing timescales 
define a maximum ``isolated'' lifetime for 
that saturated mode 
that is of interest, $\sim0.1\,t_{\rm Hubble}$ 
(Figure~\ref{fig:bar.lifetime.limit}). There 
has been substantial debate regarding the lifetime of stellar bars in disks; 
but if modes live stably in isolation for 
longer than this time, they will encounter significant cosmological 
effects including e.g.\ significant new disk growth and mergers. 
Indeed, most studies do agree that lifetimes in isolation are at least this long 
\citep[see e.g.][]{weinberg:bar.dynfric,
hernquist:bar.spheroid.interaction,
friedli:gas.bar.ssp.gradients,
athanassoula:bar.halo.growth,
kaufmann:gas.bar.evol}. 
Evolution of modes on longer timescales (e.g.\ some self-damping or buckling 
processes) should ideally be considered in a live cosmological context -- 
the time in isolation may strengthen modes against external effects, but various 
studies have found that a moderate level of 
new gas accretion or passages of new substructure 
can dramatically change mode evolution, 
both exciting and destroying bars and spiral waves 
\citep[see][]{athanassoula:bar.vs.cmc,
bournaud:gas.bar.renewal.destruction,
berentzen:bar.destruction.in.int,
berentzen:self.damping.tidal.bar.generation,
berentzen:gas.bar.interaction,foyle:two.component.disk.evol.from.bars}; 
not to mention that the presence of pre-existing strong bars may 
in turn affect these accretion/merger processes. 

Moreover, if modes live this long, their duty cycles will de-couple from 
those of their drivers. Even if, for example, all large-scale 
bars were initially driven by encounters with satellite galaxies (minor 
interactions), if the isolated lifetime were much longer than this value, 
there would be no surviving 
correlation between the {\em presence} of bars 
and such companions. There has been considerable 
observational debate regarding whether or not 
strongly-barred galaxies exhibit any strong preference for 
minor companions; certainly there are at least many such 
galaxies without close neighbors \citep[see][and
references therein]{elmegreen:bars.vs.companions,
odewahn:spiral.arms.vs.companions,
marquez:interaction.vs.spiral.dyn,marquez:interaction.vs.spiral.prop,
li:barred.gal.clustering}. This may in fact be 
because strong bars are not driven; however, it could also be consistent 
with the hypothesis that all such bars were initially driven, but are 
sufficiently long-lived. Constraints on bar lifetimes are needed to break 
the degeneracies. 

The level of cosmological dynamics also has implications for the numerical 
considerations involved in simulations of ``isolated'' systems. 
Properly following resonant self-interactions of bars may imply steep resolution 
requirements in N-body experiments 
\citep[see e.g.][]{weinberg:bar.res.1,
weinberg:bar.res.requirements,ceverino:weak.bar.res.requirements,
sellwood:weak.bar.res.requirements}. 
However, there are other properties for which increasing the 
resolution in idealized cases 
may not be a more accurate representation of reality. 
In terms of shot noise in the potential, for example, a model 
MW-like disk with $\gg 10^{6}$ particles will have potential 
fluctuations from smooth axisymmetry $\delta\phi/\phi\lesssim1\%$ over 
the spatial/timescales of interest (disk size and dynamical time). 
In cosmological simulations, although the central 
regions of halos are relatively smooth, even dark-matter only simulations 
yield comparable or larger variations in the local potential/velocity dispersion 
at e.g.\ MW-like disk effective radii \citep[see][]{zemp:via.lactea.halo.substructure}. 
Even where smooth in space, such systems are not constant in time 
(as in idealized cases) at this level over several dynamical times. 
Moreover inclusion of baryons (which are not stripped efficiently, unlike 
dark matter subhalos which are efficiently destroyed at small radii 
and so do not ``survive'' to contribute substructure inside the centers of 
halos) enhances the clumpy, minor spatial substructure. 
In the Milky Way, for example, the LMC-SMC system represents 
a real deviation from a smooth, axisymmetric potential at a level 
larger than this limit near the solar radius. 
Ideally, tracking the evolution of substructure at higher resolution 
should involve not just a larger number of particles, but cosmologically 
motivated descriptions of substructure and accretion. 

Interestingly, at all redshifts, 
we find that traditional isolated 
``secular'' evolution is most applicable around perturbations of fractional 
amplitude $\sim10\%$. 
This is a very interesting regime of parameter space: to the extent that 
it represents a fractional amplitude of substructure/accretion flows, 
it is a channel by which halos and low-mass galaxies 
gain  much of their mass \citep[e.g.][]{governato:disk.formation,
kazantzidis:mw.merger.hist.sim,stewart:mw.minor.accretion}. 
Moreover, ``pseudobulges,'' 
associated with bulge formation from secular evolution \citep[e.g.\ bar-induced inflows 
and bar buckling; see e.g.][and references therein]{oniell:bar.obs,
mayer:lsb.disk.bars,debattista:pseudobulges.a,
athanassoula:peanuts} 
appear to dominate the bulge population 
at mass ratios of similar 
amplitude \citep[$B/T \lesssim 0.1-0.2$; see][]{kuijken:pseudobulges.obs,
jogee:bar.frac.evol,kormendy.kennicutt:pseudobulge.review,
fisher:pseudobulge.sf.profile,fisher:pseudobulge.ns,weinzirl:b.t.dist}.
Suggestively, this also corresponds to typical amplitudes of 
observed strong bars 
\citep[references above and][]{eskridge:bar.freq.nir,
laurikainen:bar.strengths,sheth:bar.frac.evol,
marinova:bar.frac.vs.freq,barazza:bars.cluster.field}.

Of course, real systems exhibit more complex behavior then the simple 
scalings we derive here. Ultimately, detailed progress in modeling 
the interplay between continuous accretion of new substructure 
and cosmological driving of perturbations coupled to non-linear modes in 
galactic disks will require high-resolution $N$-body and hydrodynamic 
cosmological simulations. Some progress has begun towards modeling these 
processes in a proper cosmological context \citep[see e.g.][]{
bournaud:gas.bar.renewal.destruction,
gauthier:triggered.bar.from.substructure,
berentzen:disk.bar.in.live.halos,kaufmann:gas.bar.evol,
governato:disk.formation,
foyle:two.component.disk.evol.from.bars,
kazantzidis:mw.merger.hist.sim,
romanodiaz:substructure.driven.bars} -- these 
studies highlight a key point here, that in a large regime of parameter space 
it is difficult to disentangle ``secular'' and cosmological processes. 
Our goal here is not to derive a rigorous quantitative description of one or 
the other. However, the simple arguments here should help to 
constrain and focus the discussion of where and when (in 
realistic cosmological settings) ``isolated'' evolution is important.

\acknowledgments 
We thank Lars Hernquist and T.~J.\ Cox for helpful discussions, 
as well as Simon White and the anonymous referee for 
suggestions that greatly improved this manuscript. 
Support for PFH was provided by the Miller Institute for Basic Research 
in Science, University of California Berkeley.
DK acknowledges the support of the ITC fellowship at the 
Harvard College Observatory. 
\\

\bibliography{/Users/phopkins/Documents/lars_galaxies/papers/ms}

\begin{thebibliography}{128}
\expandafter\ifx\csname natexlab\endcsname\relax\def\natexlab#1{#1}\fi

\bibitem[{{Adams} {et~al.}(1989){Adams}, {Ruden}, \&
  {Shu}}]{adams89:eccentric.instab.in.keplerian.disks}
{Adams}, F.~C., {Ruden}, S.~P., \& {Shu}, F.~H. 1989, \apj, 347, 959

\bibitem[{{Akiyama} {et~al.}(2008){Akiyama}, {Minowa}, {Kobayashi}, {Ohta},
  {Ando}, \& {Iwata}}]{akiyama:lbg.weak.size.evol}
{Akiyama}, M., {Minowa}, Y., {Kobayashi}, N., {Ohta}, K., {Ando}, M., \&
  {Iwata}, I. 2008, \apjs, 175, 1

\bibitem[{{Athanassoula}(2002{\natexlab{a}})}]{athanassoula:bar.halo.growth}
{Athanassoula}, E. 2002{\natexlab{a}}, \apjl, 569, L83

\bibitem[{{Athanassoula}(2002{\natexlab{b}})}]{athanassoula:bar.evol.in.int}
---. 2002{\natexlab{b}}, \apss, 281, 39

\bibitem[{{Athanassoula}(2005)}]{athanassoula:peanuts}
---. 2005, \mnras, 358, 1477

\bibitem[{{Athanassoula} {et~al.}(1983){Athanassoula}, {Bienayme}, {Martinet},
  \& {Pfenniger}}]{athanassoula:bar.orbits}
{Athanassoula}, E., {Bienayme}, O., {Martinet}, L., \& {Pfenniger}, D. 1983,
  \aap, 127, 349

\bibitem[{{Athanassoula} {et~al.}(2005){Athanassoula}, {Lambert}, \&
  {Dehnen}}]{athanassoula:bar.vs.cmc}
{Athanassoula}, E., {Lambert}, J.~C., \& {Dehnen}, W. 2005, \mnras, 363, 496

\bibitem[{{Athanassoula} \&
  {Misiriotis}(2002)}]{athanassoula:bar.vs.concentration}
{Athanassoula}, E., \& {Misiriotis}, A. 2002, \mnras, 330, 35

\bibitem[{{Athanassoula} \&
  {Sellwood}(1986)}]{athanassoula.sellwood:bar.timescale}
{Athanassoula}, E., \& {Sellwood}, J.~A. 1986, \mnras, 221, 213

\bibitem[{{Barazza} {et~al.}(2008){Barazza}, {Jogee}, \&
  {Marinova}}]{barazza:bar.colors}
{Barazza}, F.~D., {Jogee}, S., \& {Marinova}, I. 2008, \apj, 675, 1194

\bibitem[{{Barazza} {et~al.}(2009)}]{barazza:bars.cluster.field}
{Barazza}, F.~D., {et~al.} 2009, \aap, in press, arXiv:0902.4080

\bibitem[{{Barnes} \& {Hernquist}(1992)}]{barneshernquist92}
{Barnes}, J.~E., \& {Hernquist}, L. 1992, \araa, 30, 705

\bibitem[{{Barteldrees} \& {Dettmar}(1994)}]{barteldrees:exp.vertical.profiles}
{Barteldrees}, A., \& {Dettmar}, R.-J. 1994, \aaps, 103, 475

\bibitem[{{Barton} {et~al.}(2007){Barton}, {Arnold}, {Zentner}, {Bullock}, \&
  {Wechsler}}]{barton:triggered.sf}
{Barton}, E.~J., {Arnold}, J.~A., {Zentner}, A.~R., {Bullock}, J.~S., \&
  {Wechsler}, R.~H. 2007, \apj, 671, 1538

\bibitem[{{Bell} \& {de Jong}(2001)}]{belldejong:tf}
{Bell}, E.~F., \& {de Jong}, R.~S. 2001, \apj, 550, 212

\bibitem[{{Benson}(2005)}]{benson:cosmo.orbits}
{Benson}, A.~J. 2005, \mnras, 358, 551

\bibitem[{{Benson} {et~al.}(2004){Benson}, {Lacey}, {Frenk}, {Baugh}, \&
  {Cole}}]{benson:heating.model}
{Benson}, A.~J., {Lacey}, C.~G., {Frenk}, C.~S., {Baugh}, C.~M., \& {Cole}, S.
  2004, \mnras, 351, 1215

\bibitem[{{Berentzen} {et~al.}(2003){Berentzen}, {Athanassoula}, {Heller}, \&
  {Fricke}}]{berentzen:bar.destruction.in.int}
{Berentzen}, I., {Athanassoula}, E., {Heller}, C.~H., \& {Fricke}, K.~J. 2003,
  \mnras, 341, 343

\bibitem[{{Berentzen} {et~al.}(2004){Berentzen}, {Athanassoula}, {Heller}, \&
  {Fricke}}]{berentzen:self.damping.tidal.bar.generation}
---. 2004, \mnras, 347, 220

\bibitem[{{Berentzen} \& {Shlosman}(2006)}]{berentzen:disk.bar.in.live.halos}
{Berentzen}, I., \& {Shlosman}, I. 2006, \apj, 648, 807

\bibitem[{{Berentzen} {et~al.}(2007){Berentzen}, {Shlosman},
  {Martinez-Valpuesta}, \& {Heller}}]{berentzen:gas.bar.interaction}
{Berentzen}, I., {Shlosman}, I., {Martinez-Valpuesta}, I., \& {Heller}, C.~H.
  2007, \apj, 666, 189

\bibitem[{{Binney} \& {Tremaine}(1987)}]{binneytremaine}
{Binney}, J., \& {Tremaine}, S. 1987, {Galactic dynamics} (Princeton, NJ,
  Princeton University Press, 1987)

\bibitem[{{Borriello} \& {Salucci}(2001)}]{borriello01}
{Borriello}, A., \& {Salucci}, P. 2001, \mnras, 323, 285

\bibitem[{{Bournaud} \& {Combes}(2002)}]{bournaud:gas.bar.renewal.destruction}
{Bournaud}, F., \& {Combes}, F. 2002, \aap, 392, 83

\bibitem[{{Boylan-Kolchin} {et~al.}(2008){Boylan-Kolchin}, {Ma}, \&
  {Quataert}}]{boylankolchin:merger.time.calibration}
{Boylan-Kolchin}, M., {Ma}, C.-P., \& {Quataert}, E. 2008, \mnras, 383, 93

\bibitem[{{Brown} {et~al.}(2007){Brown}, {Dey}, {Jannuzi}, {Brand}, {Benson},
  {Brodwin}, {Croton}, \& {Eisenhardt}}]{brown:mf.evolution}
{Brown}, M.~J.~I., {Dey}, A., {Jannuzi}, B.~T., {Brand}, K., {Benson}, A.~J.,
  {Brodwin}, M., {Croton}, D.~J., \& {Eisenhardt}, P.~R. 2007, \apj, 654, 858

\bibitem[{{Bullock} {et~al.}(2001){Bullock}, {Kolatt}, {Sigad}, {Somerville},
  {Kravtsov}, {Klypin}, {Primack}, \& {Dekel}}]{bullock:concentrations}
{Bullock}, J.~S., {Kolatt}, T.~S., {Sigad}, Y., {Somerville}, R.~S.,
  {Kravtsov}, A.~V., {Klypin}, A.~A., {Primack}, J.~R., \& {Dekel}, A. 2001,
  \mnras, 321, 559

\bibitem[{{Bundy} {et~al.}(2009){Bundy}, {Fukugita}, {Ellis}, {Targett},
  {Belli}, \& {Kodama}}]{bundy:merger.fraction.new}
{Bundy}, K., {Fukugita}, M., {Ellis}, R.~S., {Targett}, T.~A., {Belli}, S., \&
  {Kodama}, T. 2009, \apj, 697, 1369

\bibitem[{{Ceverino} \& {Klypin}(2007)}]{ceverino:weak.bar.res.requirements}
{Ceverino}, D., \& {Klypin}, A. 2007, \mnras, 379, 1155

\bibitem[{{Col{\'{\i}}n} {et~al.}(2006){Col{\'{\i}}n}, {Valenzuela}, \&
  {Klypin}}]{colin:bar.in.mw.like.halo.evol}
{Col{\'{\i}}n}, P., {Valenzuela}, O., \& {Klypin}, A. 2006, \apj, 644, 687

\bibitem[{{Combes} {et~al.}(1990){Combes}, {Debbasch}, {Friedli}, \&
  {Pfenniger}}]{combes:pseudobulges}
{Combes}, F., {Debbasch}, F., {Friedli}, D., \& {Pfenniger}, D. 1990, \aap,
  233, 82

\bibitem[{{Conroy} \& {Wechsler}(2009)}]{conroy:hod.vs.z}
{Conroy}, C., \& {Wechsler}, R.~H. 2009, \apj, 696, 620

\bibitem[{{Courteau} {et~al.}(2007){Courteau}, {Dutton}, {van den Bosch},
  {MacArthur}, {Dekel}, {McIntosh}, \& {Dale}}]{courteau:disk.scalings}
{Courteau}, S., {Dutton}, A.~A., {van den Bosch}, F.~C., {MacArthur}, L.~A.,
  {Dekel}, A., {McIntosh}, D.~H., \& {Dale}, D.~A. 2007, \apj, 671, 203

\bibitem[{{Curir} {et~al.}(2007){Curir}, {Mazzei}, \&
  {Murante}}]{curir:bar.in.cosmo.halos}
{Curir}, A., {Mazzei}, P., \& {Murante}, G. 2007, \aap, 467, 509

\bibitem[{{de Grijs} {et~al.}(1997){de Grijs}, {Peletier}, \& {van der
  Kruit}}]{degrijs:vertical.disk.profiles}
{de Grijs}, R., {Peletier}, R.~F., \& {van der Kruit}, P.~C. 1997, \aap, 327,
  966

\bibitem[{{Debattista} {et~al.}(2004){Debattista}, {Carollo}, {Mayer}, \&
  {Moore}}]{debattista:pseudobulges.a}
{Debattista}, V.~P., {Carollo}, C.~M., {Mayer}, L., \& {Moore}, B. 2004, \apjl,
  604, L93

\bibitem[{{Dekel} \& {Birnboim}(2006)}]{dekelbirnboim:mquench}
{Dekel}, A., \& {Birnboim}, Y. 2006, \mnras, 368, 2

\bibitem[{{Dekel} {et~al.}(2009){Dekel}, {Birnboim}, {Engel}, {Freundlich},
  {Goerdt}, {Mumcuoglu}, {Neistein}, {Pichon}, {Teyssier}, \&
  {Zinger}}]{dekel:cold.streams}
{Dekel}, A., {Birnboim}, Y., {Engel}, G., {Freundlich}, J., {Goerdt}, T.,
  {Mumcuoglu}, M., {Neistein}, E., {Pichon}, C., {Teyssier}, R., \& {Zinger},
  E. 2009, \nat, 457, 451

\bibitem[{{Dubinski} {et~al.}(2008){Dubinski}, {Berentzen}, \&
  {Shlosman}}]{dubinski:bar.evol.sim.tests}
{Dubinski}, J., {Berentzen}, I., \& {Shlosman}, I. 2008, \apj, in press,
  arXiv:0810.4925

\bibitem[{{Earn} \& {Sellwood}(1995)}]{earn.sellwood:95.nbody.bar.stab}
{Earn}, D.~J.~D., \& {Sellwood}, J.~A. 1995, \apj, 451, 533

\bibitem[{{Elmegreen} {et~al.}(2008){Elmegreen}, {Bournaud}, \&
  {Elmegreen}}]{elmegreen:classical.bulges.from.clumps}
{Elmegreen}, B.~G., {Bournaud}, F., \& {Elmegreen}, D.~M. 2008, \apj, 688, 67

\bibitem[{{Elmegreen} {et~al.}(1990){Elmegreen}, {Elmegreen}, \&
  {Bellin}}]{elmegreen:bars.vs.companions}
{Elmegreen}, D.~M., {Elmegreen}, B.~G., \& {Bellin}, A.~D. 1990, \apj, 364, 415

\bibitem[{{Erb} {et~al.}(2006){Erb}, {Steidel}, {Shapley}, {Pettini}, {Reddy},
  \& {Adelberger}}]{erb:lbg.gasmasses}
{Erb}, D.~K., {Steidel}, C.~C., {Shapley}, A.~E., {Pettini}, M., {Reddy},
  N.~A., \& {Adelberger}, K.~L. 2006, \apj, 646, 107

\bibitem[{{Eskridge} {et~al.}(2000)}]{eskridge:bar.freq.nir}
{Eskridge}, P.~B., {et~al.} 2000, \aj, 119, 536

\bibitem[{{Fakhouri} \& {Ma}(2008)}]{fakhouri:halo.merger.rates}
{Fakhouri}, O., \& {Ma}, C.-P. 2008, \mnras, 386, 577

\bibitem[{{Fisher}(2006)}]{fisher:pseudobulge.sf.profile}
{Fisher}, D.~B. 2006, \apjl, 642, L17

\bibitem[{{Fisher} \& {Drory}(2008)}]{fisher:pseudobulge.ns}
{Fisher}, D.~B., \& {Drory}, N. 2008, \aj, 136, 773

\bibitem[{{Flores} {et~al.}(2006){Flores}, {Hammer}, {Puech}, {Amram}, \&
  {Balkowski}}]{flores:tf.evolution}
{Flores}, H., {Hammer}, F., {Puech}, M., {Amram}, P., \& {Balkowski}, C. 2006,
  \aap, 455, 107

\bibitem[{{Fontana} {et~al.}(2006)}]{fontana:highz.mfs}
{Fontana}, A., {et~al.} 2006, \aap, 459, 745

\bibitem[{{Foyle} {et~al.}(2008){Foyle}, {Courteau}, \&
  {Thacker}}]{foyle:two.component.disk.evol.from.bars}
{Foyle}, K., {Courteau}, S., \& {Thacker}, R.~J. 2008, \mnras, 386, 1821

\bibitem[{{Friedli} \& {Benz}(1993)}]{friedli:gas.stellar.bar.evol}
{Friedli}, D., \& {Benz}, W. 1993, \aap, 268, 65

\bibitem[{{Friedli} {et~al.}(1994){Friedli}, {Benz}, \&
  {Kennicutt}}]{friedli:gas.bar.ssp.gradients}
{Friedli}, D., {Benz}, W., \& {Kennicutt}, R. 1994, \apjl, 430, L105

\bibitem[{{Gao} {et~al.}(2004){Gao}, {White}, {Jenkins}, {Stoehr}, \&
  {Springel}}]{gao:subhalo.mf}
{Gao}, L., {White}, S.~D.~M., {Jenkins}, A., {Stoehr}, F., \& {Springel}, V.
  2004, \mnras, 355, 819

\bibitem[{{Gauthier} {et~al.}(2006){Gauthier}, {Dubinski}, \&
  {Widrow}}]{gauthier:triggered.bar.from.substructure}
{Gauthier}, J.-R., {Dubinski}, J., \& {Widrow}, L.~M. 2006, \apj, 653, 1180

\bibitem[{{Genel} {et~al.}(2008)}]{genel:smg.numden.vs.mergers}
{Genel}, S., {et~al.} 2008, \apj, 688, 789

\bibitem[{{Gilmore} {et~al.}(2002){Gilmore}, {Wyse}, \&
  {Norris}}]{gilmore02:last.mw.merger.from.thick.disk}
{Gilmore}, G., {Wyse}, R.~F.~G., \& {Norris}, J.~E. 2002, \apjl, 574, L39

\bibitem[{{Governato} {et~al.}(2007){Governato}, {Willman}, {Mayer}, {Brooks},
  {Stinson}, {Valenzuela}, {Wadsley}, \& {Quinn}}]{governato:disk.formation}
{Governato}, F., {Willman}, B., {Mayer}, L., {Brooks}, A., {Stinson}, G.,
  {Valenzuela}, O., {Wadsley}, J., \& {Quinn}, T. 2007, \mnras, 374, 1479

\bibitem[{{Guo} \& {White}(2008)}]{guo:gal.growth.channels}
{Guo}, Q., \& {White}, S.~D.~M. 2008, \mnras, 384, 2

\bibitem[{{Hernquist} \& {Weinberg}(1992{\natexlab{a}})}]{hernquistweinberg92}
{Hernquist}, L., \& {Weinberg}, M.~D. 1992{\natexlab{a}}, \apj, 400, 80

\bibitem[{{Hernquist} \&
  {Weinberg}(1992{\natexlab{b}})}]{hernquist:bar.spheroid.interaction}
---. 1992{\natexlab{b}}, \apj, 400, 80

\bibitem[{{Holley-Bockelmann} {et~al.}(2005){Holley-Bockelmann}, {Weinberg}, \&
  {Katz}}]{holley:bar.halo.interaction}
{Holley-Bockelmann}, K., {Weinberg}, M., \& {Katz}, N. 2005, \mnras, 363, 991

\bibitem[{{Hopkins} {et~al.}(2008{\natexlab{a}}){Hopkins}, {Cox}, {Kere{\v s}},
  \& {Hernquist}}]{hopkins:groups.ell}
{Hopkins}, P.~F., {Cox}, T.~J., {Kere{\v s}}, D., \& {Hernquist}, L.
  2008{\natexlab{a}}, \apjs, 175, 390

\bibitem[{{Hopkins} {et~al.}(2008{\natexlab{b}}){Hopkins}, {Hernquist}, {Cox},
  {Younger}, \& {Besla}}]{hopkins:disk.heating}
{Hopkins}, P.~F., {Hernquist}, L., {Cox}, T.~J., {Younger}, J.~D., \& {Besla},
  G. 2008{\natexlab{b}}, \apj, 688, 757

\bibitem[{{Hopkins} {et~al.}(2009{\natexlab{a}}){Hopkins}, {Somerville}, {Cox},
  {Hernquist}, {Jogee}, {Kere{\v s}}, {Ma}, {Robertson}, \&
  {Stewart}}]{hopkins:disk.survival.cosmo}
{Hopkins}, P.~F., {Somerville}, R.~S., {Cox}, T.~J., {Hernquist}, L., {Jogee},
  S., {Kere{\v s}}, D., {Ma}, C.-P., {Robertson}, B., \& {Stewart}, K.
  2009{\natexlab{a}}, \mnras, 397, 802

\bibitem[{{Hopkins} {et~al.}(2009{\natexlab{b}})}]{hopkins:merger.rates}
{Hopkins}, P.~F., {et~al.} 2009{\natexlab{b}}, \mnras, in press,
  arXiv:0906.5357

\bibitem[{{Jiang} {et~al.}(2008){Jiang}, {Jing}, {Faltenbacher}, {Lin}, \&
  {Li}}]{jiang:dynfric.calibration}
{Jiang}, C.~Y., {Jing}, Y.~P., {Faltenbacher}, A., {Lin}, W.~P., \& {Li}, C.
  2008, \apj, 675, 1095

\bibitem[{{Jogee} {et~al.}(2004)}]{jogee:bar.frac.evol}
{Jogee}, S., {et~al.} 2004, \apjl, 615, L105

\bibitem[{{Kassin} {et~al.}(2007)}]{kassin:tf.evolution}
{Kassin}, S.~A., {et~al.} 2007, \apjl, 660, L35

\bibitem[{{Kaufmann} {et~al.}(2007){Kaufmann}, {Mayer}, {Wadsley}, {Stadel}, \&
  {Moore}}]{kaufmann:gas.bar.evol}
{Kaufmann}, T., {Mayer}, L., {Wadsley}, J., {Stadel}, J., \& {Moore}, B. 2007,
  \mnras, 375, 53

\bibitem[{{Kazantzidis} {et~al.}(2008){Kazantzidis}, {Bullock}, {Zentner},
  {Kravtsov}, \& {Moustakas}}]{kazantzidis:mw.merger.hist.sim}
{Kazantzidis}, S., {Bullock}, J.~S., {Zentner}, A.~R., {Kravtsov}, A.~V., \&
  {Moustakas}, L.~A. 2008, \apj, 688, 254

\bibitem[{{Kere{\v s}} {et~al.}(2009){Kere{\v s}}, {Katz}, {Fardal},
  {Dav{\'e}}, \& {Weinberg}}]{keres:cooling.revised}
{Kere{\v s}}, D., {Katz}, N., {Fardal}, M., {Dav{\'e}}, R., \& {Weinberg},
  D.~H. 2009, \mnras, 395, 160

\bibitem[{{Kere{\v s}} {et~al.}(2005){Kere{\v s}}, {Katz}, {Weinberg}, \&
  {Dav{\'e}}}]{keres:hot.halos}
{Kere{\v s}}, D., {Katz}, N., {Weinberg}, D.~H., \& {Dav{\'e}}, R. 2005,
  \mnras, 363, 2

\bibitem[{{Khochfar} \& {Burkert}(2006)}]{khochfar:cosmo.orbits}
{Khochfar}, S., \& {Burkert}, A. 2006, \aap, 445, 403

\bibitem[{{Kitzbichler} \&
  {White}(2008)}]{kitzbichler:mgr.rate.pair.calibration}
{Kitzbichler}, M.~G., \& {White}, S.~D.~M. 2008, \mnras, 391, 1489

\bibitem[{{Klypin} {et~al.}(2008){Klypin}, {Valenzuela}, {Colin}, \&
  {Quinn}}]{klypin:bar.dynamics.vs.thickness}
{Klypin}, A., {Valenzuela}, O., {Colin}, P., \& {Quinn}, T. 2008, \mnras, in
  press, arXiv:0808.3422

\bibitem[{{Komatsu} {et~al.}(2009)}]{komatsu:wmap5}
{Komatsu}, E., {et~al.} 2009, \apjs, 180, 330

\bibitem[{{Kormendy} \&
  {Kennicutt}(2004)}]{kormendy.kennicutt:pseudobulge.review}
{Kormendy}, J., \& {Kennicutt}, Jr., R.~C. 2004, \araa, 42, 603

\bibitem[{{Kuijken} \& {Merrifield}(1995)}]{kuijken:pseudobulges.obs}
{Kuijken}, K., \& {Merrifield}, M.~R. 1995, \apjl, 443, L13

\bibitem[{{Laurikainen} {et~al.}(2002){Laurikainen}, {Salo}, \&
  {Rautiainen}}]{laurikainen:bar.strengths}
{Laurikainen}, E., {Salo}, H., \& {Rautiainen}, P. 2002, \mnras, 331, 880

\bibitem[{{Li} {et~al.}(2009){Li}, {Gadotti}, {Mao}, \&
  {Kauffmann}}]{li:barred.gal.clustering}
{Li}, C., {Gadotti}, D.~A., {Mao}, S., \& {Kauffmann}, G. 2009, \mnras, in
  press, arXiv:0902.1175

\bibitem[{{Maller} {et~al.}(2006){Maller}, {Katz}, {Kere{\v s}}, {Dav{\'e}}, \&
  {Weinberg}}]{maller:sph.merger.rates}
{Maller}, A.~H., {Katz}, N., {Kere{\v s}}, D., {Dav{\'e}}, R., \& {Weinberg},
  D.~H. 2006, \apj, 647, 763

\bibitem[{{Marinova} \& {Jogee}(2007)}]{marinova:bar.frac.vs.freq}
{Marinova}, I., \& {Jogee}, S. 2007, \apj, 659, 1176

\bibitem[{{Marquez} \& {Moles}(1996)}]{marquez:interaction.vs.spiral.prop}
{Marquez}, I., \& {Moles}, M. 1996, \aaps, 120, 1

\bibitem[{{Martinez-Valpuesta} {et~al.}(2006){Martinez-Valpuesta}, {Shlosman},
  \& {Heller}}]{martinezvalpuesta:recurrent.buckling}
{Martinez-Valpuesta}, I., {Shlosman}, I., \& {Heller}, C. 2006, \apj, 637, 214

\bibitem[{{Mayer} \& {Wadsley}(2004)}]{mayer:lsb.disk.bars}
{Mayer}, L., \& {Wadsley}, J. 2004, \mnras, 347, 277

\bibitem[{{McGaugh}(2005)}]{mcgaugh:tf}
{McGaugh}, S.~S. 2005, \apj, 632, 859

\bibitem[{{Men{\'e}ndez-Delmestre} {et~al.}(2007){Men{\'e}ndez-Delmestre},
  {Sheth}, {Schinnerer}, {Jarrett}, \& {Scoville}}]{menendez:2mass.bars}
{Men{\'e}ndez-Delmestre}, K., {Sheth}, K., {Schinnerer}, E., {Jarrett}, T.~H.,
  \& {Scoville}, N.~Z. 2007, \apj, 657, 790

\bibitem[{{Mestel}(1963)}]{mestel:disk.profile}
{Mestel}, L. 1963, \mnras, 126, 553

\bibitem[{{Mihos} {et~al.}(1997){Mihos}, {McGaugh}, \& {de
  Blok}}]{mihos:lsb.gal.stable.vs.bars}
{Mihos}, J.~C., {McGaugh}, S.~S., \& {de Blok}, W.~J.~G. 1997, \apjl, 477, L79+

\bibitem[{{Mo} {et~al.}(1998){Mo}, {Mao}, \& {White}}]{momauwhite:disks}
{Mo}, H.~J., {Mao}, S., \& {White}, S.~D.~M. 1998, \mnras, 295, 319

\bibitem[{{Moles} {et~al.}(1995){Moles}, {Marquez}, \&
  {Perez}}]{marquez:interaction.vs.spiral.dyn}
{Moles}, M., {Marquez}, I., \& {Perez}, E. 1995, \apj, 438, 604

\bibitem[{{Narayan} {et~al.}(1987){Narayan}, {Goldreich}, \&
  {Goodman}}]{narayan:87.bar.tscale.shearing.sheet}
{Narayan}, R., {Goldreich}, P., \& {Goodman}, J. 1987, \mnras, 228, 1

\bibitem[{{Noeske} {et~al.}(2007)}]{noeske:sfh}
{Noeske}, K.~G., {et~al.} 2007, \apjl, 660, L47

\bibitem[{{Odewahn}(1994)}]{odewahn:spiral.arms.vs.companions}
{Odewahn}, S.~C. 1994, \aj, 107, 1320

\bibitem[{{O'Neill} \& {Dubinski}(2003)}]{oniell:bar.obs}
{O'Neill}, J.~K., \& {Dubinski}, J. 2003, \mnras, 346, 251

\bibitem[{{Parry} {et~al.}(2008){Parry}, {Eke}, \&
  {Frenk}}]{parry:sam.merger.vs.morph}
{Parry}, O.~H., {Eke}, V.~R., \& {Frenk}, C.~S. 2008, \mnras, in press,
  arXiv:0806.4189

\bibitem[{{Patsis} \& {Athanassoula}(2000)}]{patsis:gas.flow.in.bars}
{Patsis}, P.~A., \& {Athanassoula}, E. 2000, \aap, 358, 45

\bibitem[{{P{\'e}rez-Gonz{\'a}lez}
  {et~al.}(2008)}]{perezgonzalez:mf.compilation}
{P{\'e}rez-Gonz{\'a}lez}, P.~G., {et~al.} 2008, \apj, 675, 234

\bibitem[{{Persic} {et~al.}(1996){Persic}, {Salucci}, \& {Stel}}]{persic96}
{Persic}, M., {Salucci}, P., \& {Stel}, F. 1996, \mnras, 281, 27

\bibitem[{{Pfenniger}(1984)}]{pfenniger:bar.dynamics}
{Pfenniger}, D. 1984, \aap, 134, 373

\bibitem[{{Romano-Diaz} {et~al.}(2008){Romano-Diaz}, {Shlosman}, {Heller}, \&
  {Hoffman}}]{romanodiaz:substructure.driven.bars}
{Romano-Diaz}, E., {Shlosman}, I., {Heller}, C., \& {Hoffman}, Y. 2008, \apj,
  in press arXiv:0809.2785 [astro-ph]

\bibitem[{{Schwarz}(1981)}]{schwarz:disk-bar}
{Schwarz}, M.~P. 1981, \apj, 247, 77

\bibitem[{{Sellwood}(2008)}]{sellwood:weak.bar.res.requirements}
{Sellwood}, J.~A. 2008, \apj, 679, 379

\bibitem[{{Shapiro} {et~al.}(2008)}]{shapiro:highz.kinematics}
{Shapiro}, K.~L., {et~al.} 2008, \apj, 682, 231

\bibitem[{{Shen} {et~al.}(2003){Shen}, {Mo}, {White}, {Blanton}, {Kauffmann},
  {Voges}, {Brinkmann}, \& {Csabai}}]{shen:size.mass}
{Shen}, S., {Mo}, H.~J., {White}, S.~D.~M., {Blanton}, M.~R., {Kauffmann}, G.,
  {Voges}, W., {Brinkmann}, J., \& {Csabai}, I. 2003, \mnras, 343, 978

\bibitem[{{Sheth} {et~al.}(2003){Sheth}, {Regan}, {Scoville}, \&
  {Strubbe}}]{sheth:bar.frac.evol}
{Sheth}, K., {Regan}, M.~W., {Scoville}, N.~Z., \& {Strubbe}, L.~E. 2003,
  \apjl, 592, L13

\bibitem[{{Sheth} {et~al.}(2008)}]{sheth:bar.evol.cosmos}
{Sheth}, K., {et~al.} 2008, \apj, 675, 1141

\bibitem[{{Shu} {et~al.}(1990){Shu}, {Tremaine}, {Adams}, \&
  {Ruden}}]{shu:gas.disk.bar.tscale}
{Shu}, F.~H., {Tremaine}, S., {Adams}, F.~C., \& {Ruden}, S.~P. 1990, \apj,
  358, 495

\bibitem[{{Somerville} {et~al.}(2008)}]{somerville:disk.size.evol}
{Somerville}, R.~S., {et~al.} 2008, \apj, 672, 776

\bibitem[{{Stewart} {et~al.}(2008{\natexlab{a}}){Stewart}, {Bullock}, {Barton},
  \& {Wechsler}}]{stewart:merger.rates}
{Stewart}, K.~R., {Bullock}, J.~S., {Barton}, E.~J., \& {Wechsler}, R.~H.
  2008{\natexlab{a}}, \apj, in press, arXiv:0811.1218 [astro-ph]

\bibitem[{{Stewart} {et~al.}(2009){Stewart}, {Bullock}, {Wechsler}, \&
  {Maller}}]{stewart:disk.survival.vs.mergerrates}
{Stewart}, K.~R., {Bullock}, J.~S., {Wechsler}, R.~H., \& {Maller}, A.~H. 2009,
  \apj, in press, arXiv:0901.4336

\bibitem[{{Stewart} {et~al.}(2008{\natexlab{b}}){Stewart}, {Bullock},
  {Wechsler}, {Maller}, \& {Zentner}}]{stewart:mw.minor.accretion}
{Stewart}, K.~R., {Bullock}, J.~S., {Wechsler}, R.~H., {Maller}, A.~H., \&
  {Zentner}, A.~R. 2008{\natexlab{b}}, \apj, 683, 597

\bibitem[{{Taylor} \& {Babul}(2004)}]{taylor:substructure.evolution}
{Taylor}, J.~E., \& {Babul}, A. 2004, \mnras, 348, 811

\bibitem[{{Toft} {et~al.}(2007)}]{toft:z2.sizes.vs.sfr}
{Toft}, S., {et~al.} 2007, \apj, 671, 285

\bibitem[{{Toomre}(1981)}]{toomre:spiral.mode.growth}
{Toomre}, A. 1981, in Structure and Evolution of Normal Galaxies (Cambridge
  University Press, NY), ed. S.~M. {Fall} \& D.~{Lynden-Bell}, 111--136

\bibitem[{{Trujillo} {et~al.}(2006)}]{trujillo:size.evolution}
{Trujillo}, I., {et~al.} 2006, \apj, 650, 18

\bibitem[{{van Starkenburg} {et~al.}(2008){van Starkenburg}, {van der Werf},
  {Franx}, {Labb{\'e}}, {Rudnick}, \&
  {Wuyts}}]{vanstarkenburg:z2.disk.dynamics}
{van Starkenburg}, L., {van der Werf}, P.~P., {Franx}, M., {Labb{\'e}}, I.,
  {Rudnick}, G., \& {Wuyts}, S. 2008, \aap, 488, 99

\bibitem[{{Velazquez} \& {White}(1999)}]{velazquezwhite:disk.heating}
{Velazquez}, H., \& {White}, S.~D.~M. 1999, \mnras, 304, 254

\bibitem[{{Wechsler} {et~al.}(2002){Wechsler}, {Bullock}, {Primack},
  {Kravtsov}, \& {Dekel}}]{wechsler:concentration}
{Wechsler}, R.~H., {Bullock}, J.~S., {Primack}, J.~R., {Kravtsov}, A.~V., \&
  {Dekel}, A. 2002, \apj, 568, 52

\bibitem[{{Weinberg}(1985)}]{weinberg:bar.dynfric}
{Weinberg}, M.~D. 1985, \mnras, 213, 451

\bibitem[{{Weinberg} \& {Katz}(2007{\natexlab{a}})}]{weinberg:bar.res.1}
{Weinberg}, M.~D., \& {Katz}, N. 2007{\natexlab{a}}, \mnras, 375, 425

\bibitem[{{Weinberg} \&
  {Katz}(2007{\natexlab{b}})}]{weinberg:bar.res.requirements}
---. 2007{\natexlab{b}}, \mnras, 375, 460

\bibitem[{{Weinzirl} {et~al.}(2009){Weinzirl}, {Jogee}, {Khochfar}, {Burkert},
  \& {Kormendy}}]{weinzirl:b.t.dist}
{Weinzirl}, T., {Jogee}, S., {Khochfar}, S., {Burkert}, A., \& {Kormendy}, J.
  2009, \apj, 696, 411

\bibitem[{{Wetzel} {et~al.}(2009){Wetzel}, {Cohn}, \&
  {White}}]{wetzel:mgr.rate.subhalos}
{Wetzel}, A.~R., {Cohn}, J.~D., \& {White}, M. 2009, \mnras, 395, 1376

\bibitem[{{Woods} \& {Geller}(2007)}]{woods:minor.mergers}
{Woods}, D.~F., \& {Geller}, M.~J. 2007, \aj, 134, 527

\bibitem[{{Woods} {et~al.}(2006){Woods}, {Geller}, \&
  {Barton}}]{woods:tidal.triggering}
{Woods}, D.~F., {Geller}, M.~J., \& {Barton}, E.~J. 2006, \aj, 132, 197

\bibitem[{{Wyse} {et~al.}(2006){Wyse}, {Gilmore}, {Norris}, {Wilkinson},
  {Kleyna}, {Koch}, {Evans}, \& {Grebel}}]{wyse06:thick.disk.stars}
{Wyse}, R.~F.~G., {Gilmore}, G., {Norris}, J.~E., {Wilkinson}, M.~I., {Kleyna},
  J.~T., {Koch}, A., {Evans}, N.~W., \& {Grebel}, E.~K. 2006, \apjl, 639, L13

\bibitem[{{Zemp} {et~al.}(2008){Zemp}, {Diemand}, {Kuhlen}, {Madau}, {Moore},
  {Potter}, {Stadel}, \& {Widrow}}]{zemp:via.lactea.halo.substructure}
{Zemp}, M., {Diemand}, J., {Kuhlen}, M., {Madau}, P., {Moore}, B., {Potter},
  D., {Stadel}, J., \& {Widrow}, L. 2008, \mnras, in press, arXiv:0812.2033

\end{thebibliography}

\end{document}